\documentclass[aps,prd,twocolumn,nofootinbib,superscriptaddress,floatfix]{revtex4-2}

\usepackage{amssymb}
\usepackage{amsmath}
\usepackage{bm}
\usepackage{graphicx}
\usepackage{dcolumn}
\usepackage{xcolor}
\definecolor{crimsonglory}{rgb}{0.75,0.0,0.2}
\usepackage[colorlinks=true,linkcolor=crimsonglory,urlcolor=blue,citecolor=blue]{hyperref}

\newcommand{\MBH}{M_{\rm BH}}
\newcommand{\MADM}{M_{\rm ADM}}
\newcommand{\rPS}{r_{\rm PS}}
\newcommand{\rSh}{r_{\rm Sh}}
\newcommand{\PhiH}{\Phi_{\rm H}}
\newcommand{\Msun}{M_\odot}

\begin{document}
	
	\title{Black Hole Shadows and Quasinormal Modes with Dark Matter at the Horizon}
	
	\author{Armin Sadeghi}
	\email{armin.sadeghi.h@gmail.com}
	\affiliation{Department of Physics, K.N. Toosi University of Technology, P.O. Box 15875-4416, Tehran, Iran}
	
	\author{Javad T. Firouzjaee}
	\email{firouzjaee@kntu.ac.ir}
	\affiliation{Department of Physics, K.N. Toosi University of Technology, P.O. Box 15875-4416, Tehran, Iran}
	
	\author{Saeed Fakhry}
	\email{saeed.fakhry@uv.es}
	\affiliation{Departamento de Astronom\'ia y Astrof\'isica, Universitat de Val\`encia, Avinguda Vicent Andr\'es Estell\'es 19, 46100, Burjassot (Val\`encia), Spain}
	
	\date{\today}
	
	\begin{abstract}
		In this work, we study the effect of a surrounding dark matter halo on two horizon-scale observables of a black hole: its shadow and quasinormal-mode spectrum. We consider a Schwarzschild black hole embedded in a Hernquist halo and construct a one-parameter family of solutions that preserve the total halo mass while varying how the halo density vanishes at the horizon. By closing the Einstein equations with an Einstein-cluster stress tensor, we show that horizon regularity requires the density to vanish at least linearly. Our results also indicate that the dominant energy condition is violated in a thin layer outside the horizon, highlighting the limitations of the Einstein-cluster description within the photon sphere. We demonstrate that, at fixed black hole mass, the halo always enlarges the shadow, with the effect becoming more pronounced for more massive and compact halos. Comparison with Event Horizon Telescope observations of Sgr\,A$^*$ and M87$^*$ shows that only sufficiently compact halos can be constrained, with the inferred bounds depending sensitively on whether the observed mass is identified with the black hole mass or the total mass of the system. We further show that the halo lowers both the oscillation frequencies and damping rates of electromagnetic and axial gravitational quasinormal modes, leading to higher quality factors. These spectral shifts are quantitatively explained by changes in the photon-sphere properties through the eikonal correspondence. Since the effects scale with the ratio of the halo mass to its size, they are observable only for dark matter concentrated within a few tens of gravitational radii of the black hole, which is the regime considered here.
	\end{abstract}
	\maketitle
	
	\section{Introduction}\label{sec:intro}
	
	Black holes (BHs) are now probed directly in two complementary ways. Gravitational waves (GWs) from merging compact binaries, first detected from a binary BH merger by the LIGO and Virgo collaborations \cite{LIGOScientific:2016aoc} and now observed in large numbers \cite{KAGRA:2021vkt}, give access to the strong-field dynamics of spacetime. At the same time, the Event Horizon Telescope (EHT) has produced horizon-scale images of M87$^*$ and Sgr\,A$^*$, whose central brightness depressions are interpreted as BH shadows \cite{EventHorizonTelescope:2019dse,EventHorizonTelescope:2022wkp}. Both channels are sensitive to the spacetime geometry within a few gravitational radii of the horizon, and therefore, in principle, to any matter that resides there.
	
	After a merger, the remnant BH relaxes through a ringdown phase dominated by quasinormal modes (QNMs), i.e., damped oscillations with complex frequencies $\omega=\omega_R+i\omega_I$, where $\omega_R$ is the oscillation frequency and $|\omega_I|$ the damping rate. They are defined by purely ingoing boundary conditions at the horizon and purely outgoing ones at spatial infinity \cite{Kokkotas:1999bd,Berti:2009kk,Konoplya:2011qq}. For a static, spherically symmetric background, perturbations can be decomposed into spherical harmonics and reduced to a Schr\"odinger-like equation \cite{Regge:1957td}. When the effective potential has a single, smooth maximum, the Wentzel-Kramers-Brillouin (WKB) approximation provides a semianalytic determination of the QNMs. It was introduced at leading order by Schutz and Will \cite{Schutz:1985km}, extended to third order by Iyer and Will \cite{Iyer:1986np} and to sixth order by Konoplya \cite{Konoplya:2003ii}, and has later been improved by Pad\'e resummation and by going to even higher orders \cite{Matyjasek:2017psv,Konoplya:2019hlu}.
	
	Dark matter (DM) is inferred from its gravitational effects on visible matter on galactic and cosmological scales \cite{Clowe:2006eq,Bertone:2004pz}. In the cold DM paradigm, galaxies are embedded in extended halos with cuspy density profiles such as the Navarro-Frenk-White (NFW) profile \cite{Navarro:1995iw,Navarro:1996gj}, and the adiabatic growth of a central BH may further steepen the DM distribution into a ``spike'' \cite{Gondolo:1999ef,Sadeghian:2013laa}. The DM environment is not a passive background for the compact objects it hosts. The halo density profile, concentration parameter, and mass function, the collapse model of the halo, and the nature of the DM itself can strongly modify the dynamics of compact objects in halos and, in particular, the predicted merger rates of binary compact objects, see, e.g., our previous works \cite{Fakhry:2020plg,Fakhry:2021tzk,Fakhry:2022zum,Fakhry:2023dpm, Fakhry:2022pbhns,Fakhry:2024pbhns, Fakhry:2023frg,Fakhry:2024pbhns}. These effects are particularly pronounced in the dense DM spikes expected around supermassive BHs, where the merger rates of compact binaries depend sensitively on the spike profile and on the underlying gravitational theory, see, e.g., \cite{Fakhry:2023spike,Fakhry:2024spikeMG}. This sensitivity suggests that the DM environment may also leave imprints on the strong-field properties of the central BH itself. Whether, and how much, DM is present within a few gravitational radii of a supermassive BH, however, remains an open question that horizon-scale observations may help to address.
	
	Light passing close to a BH is strongly deflected. Outside the horizon there is a radius, the photon sphere, at which photons can move on unstable circular orbits. Photons arriving from infinity with impact parameter below a critical value $b_c$ are captured, whereas those with $b>b_c$ are deflected and may reach a distant observer \cite{Falcke:1999pj,Perlick:2021aok}. The region of the observer's sky bounded by the critical rays appears dark and constitutes the BH shadow. Because the photon sphere and the critical rays probe the region outside the horizon, a DM distribution around the BH modifies the shadow, and the comparison of predicted shadow sizes with EHT measurements constrains such models \cite{Bambi:2019tjh,Vagnozzi:2022moj}. In the eikonal limit, the real part and the damping rate of the QNMs are determined by the angular velocity and the Lyapunov exponent of the photon sphere \cite{Cardoso:2008bp}, so the shadow and the ringdown are two manifestations of the same geometric structure.
	
	BHs in galactic environments have consequently received considerable attention. Their impact on GW generation and propagation was studied in Refs.~\cite{Cardoso:2021wlq,Cardoso:2022whc,Figueiredo:2023gas}. Spacetimes of BHs surrounded by DM halos were constructed in Refs.~\cite{Xu:2018wow,Cardoso:2021wlq,Konoplya:2022hbl,Shen:2023erj}; their QNMs and shadows were investigated in Refs.~\cite{Jusufi:2019nrn,Konoplya:2021ube,Jusufi:2022jxu,Pezzella:2024tkf}; polar perturbations, which couple to the matter sector, were analyzed in Ref.~\cite{Speeney:2024mas}; energy conditions and sound speeds were discussed in Ref.~\cite{Datta:2023zmd}; and the Einstein cluster as a model for a spiky central DM distribution was examined in Ref.~\cite{Maeda:2024tsg}. A widely used solution is that of Cardoso \textit{et al.}~\cite{Cardoso:2021wlq}, who embedded a Schwarzschild BH in a Hernquist halo using the Einstein-cluster closure \cite{Einstein:1939ms} and obtained the metric in closed form. In that construction the halo density vanishes linearly in $r-2\MBH$ at the horizon. This specific choice is not dictated by the Hernquist profile, which describes the halo on scales $r\gg\MBH$, and it is natural to ask which properties of the solution and of its observables depend on how the density is switched off near the horizon.
	
In this paper, we address this question by constructing a one-parameter family of Schwarzschild BHs embedded in a Hernquist halo, labeled by an exponent $\sigma>0$ that sets the near-horizon behavior of the density, $\rho\propto(r-2\MBH)^\sigma$, while keeping the halo mass, and hence the Arnowitt-Deser-Misner (ADM) mass $\MBH+M$, fixed; the solution of Ref.~\cite{Cardoso:2021wlq} corresponds to $\sigma=1$. We derive the exact metric and show that the horizon is regular if and only if $\sigma\ge1$ (Sec.~\ref{sec:model}), and that the region in which the Einstein-cluster closure loses its particle interpretation lies inside the photon sphere (Sec.~\ref{sec:EC}). We then prove that, at fixed BH mass, the halo always enlarges the shadow, and confront this result with the EHT observations of Sgr\,A$^*$ and M87$^*$, with particular attention to which mass the observed shadow size should be compared with (Sec.~\ref{sec:shadow}). Finally, we compute the QNMs of electromagnetic and axial gravitational perturbations with the sixth-order WKB method and show that their halo-induced shifts are explained by the eikonal correspondence with the photon sphere, even for the lowest multipoles (Sec.~\ref{sec:QNM}). Throughout, we stress that the direction of the effects depends on the mass unit, since in units of the total mass they are reversed, and that for galactic halos all effects are suppressed by $M/a\sim10^{-5}$, so our bounds apply only to compact DM concentrations. Section~\ref{sec:Conc} summarizes our conclusions, and Appendices~\ref{app:family} and~\ref{app:numerics} contain the derivation of the general family and the numerical validation. We use the signature $(-,+,+,+)$ and units $G=c=1$; $\MBH$ is the BH mass, $M$ the total halo mass, $a$ the halo scale radius, $f(r)\equiv1-2\MBH/r$, and $\MADM=\MBH+M$. Unless otherwise stated, numerical results are given for $a=\MBH$ and in units of $\MBH$.
	
	\section{Black hole embedded in a Hernquist halo}\label{sec:model}
	
	\subsection{Field equations and Einstein-cluster closure}\label{sec:fieldeqs}
	
	We consider the static, spherically symmetric line element
	\begin{equation}
		ds^2=-e^{\Phi(r)}dt^2+\left(1-\frac{2m(r)}{r}\right)^{-1}dr^2+r^2d\Omega^2,\label{genmet}
	\end{equation}
	with $d\Omega^2=d\theta^2+\sin^2\theta\,d\phi^2$, where $m(r)$ is the Misner-Sharp mass function \cite{Misner:1964je}. For an anisotropic fluid, $T^\mu{}_\nu=\mathrm{diag}(-\rho,p_r,p_t,p_t)$, the Einstein equations $G^\mu{}_\nu=8\pi T^\mu{}_\nu$ read
	\begin{align}
		G^t{}_t&=-\frac{2m'}{r^2}=-8\pi\rho,\label{Gtt}\\
		G^r{}_r&=\left(1-\frac{2m}{r}\right)\frac{\Phi'}{r}-\frac{2m}{r^3}=8\pi p_r,\label{Grr}\\
		G^\theta{}_\theta&=\frac{m}{r^3}-\frac{m'}{r^2}+\frac14\left(2\Phi''+\Phi'^2\right)\left(1-\frac{2m}{r}\right)\nonumber\\
		&\quad-\left(m'-1+\frac{m}{r}\right)\frac{\Phi'}{2r}=8\pi p_t,\label{Gthth}
	\end{align}
	where a prime denotes $d/dr$. Equation~\eqref{Gtt} gives
	\begin{equation}
		m'(r)=4\pi r^2\rho(r).\label{mprime}
	\end{equation}
	The system is underdetermined, and we close it with the Einstein-cluster ansatz $p_r=0$ \cite{Einstein:1939ms,Cardoso:2021wlq,Cardoso:2022whc,Figueiredo:2023gas}, which describes a spherical swarm of gravitating particles on circular geodesics with all orientations of the orbital planes. Setting $p_r=0$ in Eq.~\eqref{Grr} yields
	\begin{equation}
		\Phi'(r)=\frac{2m(r)}{r\left[r-2m(r)\right]},\label{eq:Phi_prime}
	\end{equation}
	and Eq.~\eqref{Gthth}, or equivalently the conservation equation $\nabla_\mu T^{\mu}{}_{r}=0$, gives the tangential pressure and the Ricci scalar,
	\begin{align}
		p_t&=\frac{m\,\rho}{2(r-2m)},\label{eq:pt}\\
		R&=8\pi(\rho-2p_t)=\frac{2m'(r-3m)}{r^2(r-2m)}.\label{Ricci}
	\end{align}
	For later use we also give the Kretschmann scalar, which for the closure \eqref{eq:Phi_prime} takes the compact form
	\begin{align}
		K\equiv R_{\mu\nu\alpha\beta}R^{\mu\nu\alpha\beta}=\;&\frac{48m^2}{r^6}-\frac{16\,m\,m'(2r-3m)}{r^5(r-2m)}\nonumber\\
		&+\frac{4m'^2(3r^2-10rm+9m^2)}{r^4(r-2m)^2}.\label{Kretsch}
	\end{align}
	Thus, once $\rho(r)$ is specified, the spacetime is completely determined by $m(r)$.
	
	\subsection{Halo density profiles}\label{sec:profiles}
	
	A DM halo is characterized by its density profile $\rho(r)$. A necessary condition for embedding it in an asymptotically flat spacetime with finite ADM mass is that the total mass be finite. If $\rho\propto r^{-\beta}$ at large $r$, the mass $m(r)\propto\int^r r'^{2-\beta}dr'$ converges if and only if
	\begin{equation}
		\beta>3.\label{mpropbeta}
	\end{equation}
	A commonly used family of double power-law profiles is \cite{Dehnen:1993uh,Zhao:1995cp,Kravtsov:1997dp}
	\begin{equation}
		\rho(r)=\rho_s\left(\frac{r}{a}\right)^{-\gamma}\left[1+\left(\frac{r}{a}\right)^{\alpha}\right]^{-(\beta-\gamma)/\alpha},\label{rhoGen}
	\end{equation}
	where $\rho_s$ is a characteristic density, $a$ a scale radius, and $\alpha$ controls the sharpness of the transition between the inner slope $\gamma$ and the outer slope $\beta$. The NFW profile, $(\alpha,\beta,\gamma)=(1,3,1)$ \cite{Navarro:1995iw,Navarro:1996gj}, violates Eq.~\eqref{mpropbeta}: its mass diverges logarithmically and must be truncated. The Hernquist profile \cite{Hernquist:1990be}, introduced as an analytic model of spherical galaxies and bulges and as an alternative to the Jaffe model \cite{Jaffe:1983iv}, corresponds to $(\alpha,\beta,\gamma)=(1,4,1)$ with $\rho_s=M/(2\pi a^3)$:
	\begin{equation}
		\rho_{\rm H}(r)=\frac{M}{2\pi}\frac{a}{r(r+a)^3},\qquad m_{\rm H}(r)=\frac{M\,r^2}{(r+a)^2},\label{rhoH}
	\end{equation}
	where $M$ is the total mass; half of it is enclosed within $r=(1+\sqrt2)\,a$. Since $\beta=4$, the Hernquist profile satisfies Eq.~\eqref{mpropbeta} and requires no truncation at large radii. It shares the inner cusp $\rho\propto r^{-1}$ of the NFW profile, which makes it a convenient finite-mass proxy for cold DM halos.
	
	\subsection{A one-parameter family of horizon truncations}\label{sec:embedded}
	
	We require that the halo does not modify the BH horizon and interior, and impose the following conditions:
	\begin{enumerate}
		\item[(C1)] $\rho(r)=0$ for $r\le2\MBH$, and $\rho(r)\to0$ as $r\to2\MBH^+$;
		\item[(C2)] the mass function satisfies
		\begin{equation}
			m(2\MBH)=\MBH,\qquad \lim_{r\to\infty}m(r)=\MBH+M,\label{cond2}
		\end{equation}
		so that the horizon is at $r=2\MBH$ and the ADM mass is the sum of the BH and halo masses;
		\item[(C3)] the Hernquist profile is recovered for $\MBH\to0$, and the Schwarzschild solution for $M\to0$.
	\end{enumerate}
	A simple way to satisfy these conditions is to multiply the Hernquist density by a power of the Schwarzschild redshift factor $f=1-2\MBH/r$ and by a function $A_\sigma(r)$ that fixes the normalization:
	\begin{equation}
		\rho_\sigma(r)=
		\begin{cases}
			\dfrac{aM}{2\pi r(a+r)^3}\,A_\sigma(r)\,f(r)^{\sigma}, & r>2\MBH,\\[2mm]
			0, & r\le2\MBH.
		\end{cases}\label{rhosigma}
	\end{equation}
	The exponent $\sigma>0$ controls how fast the density is switched off at the horizon, $\rho_\sigma\propto(r-2\MBH)^\sigma$. As shown in Appendix~\ref{app:family}, integrating Eq.~\eqref{mprime} with $m(2\MBH)=\MBH$, condition (C2) holds for every $\sigma>0$ provided
	\begin{align}
		A_\sigma(r)&=1+(\sigma-1)\frac{\MBH}{r}+(\sigma+1)\frac{\MBH}{a},\label{Asigma}\\
		m_\sigma(r)&=\MBH+\frac{M\,r^2}{(a+r)^2}\,f(r)^{\sigma+1}\qquad(r\ge2\MBH).\label{msigma}
	\end{align}
	Since $A_\sigma>0$ for $r\ge2\MBH$ and $\sigma>0$, the density is non-negative outside the horizon, and $m_\sigma$ increases monotonically from $\MBH$ to $\MBH+M$. For $\MBH\to0$ one has $A_\sigma\to1$ and $f\to1$, recovering Eq.~\eqref{rhoH}. In this work we focus on three representative members,
	\begin{align}
		\rho_{1/2}&=\frac{aM}{2\pi r(a+r)^3}\left[1+\frac{\MBH}{2r}\left(\frac{3r}{a}-1\right)\right]\nonumber\\
		&\quad\times\sqrt{1-\frac{2\MBH}{r}},\label{rho1/2}\\
		\rho_{1}&=\frac{M(a+2\MBH)}{2\pi r(r+a)^3}\left(1-\frac{2\MBH}{r}\right),\label{rho1}\\
		\rho_{3/2}&=\frac{aM}{2\pi r(a+r)^3}\left(1+\frac{\MBH}{2r}+\frac{5\MBH}{2a}\right)\nonumber\\
		&\quad\times\left(1-\frac{2\MBH}{r}\right)^{3/2},\label{rho3/2}
	\end{align}
	for $r>2\MBH$. The profile $\rho_1$ is the one introduced by Cardoso \textit{et al.}~\cite{Cardoso:2021wlq} and subsequently used, e.g., in Refs.~\cite{Konoplya:2021ube,Pezzella:2024tkf}.
	
\begin{figure}[t]
		\includegraphics[width=\columnwidth]{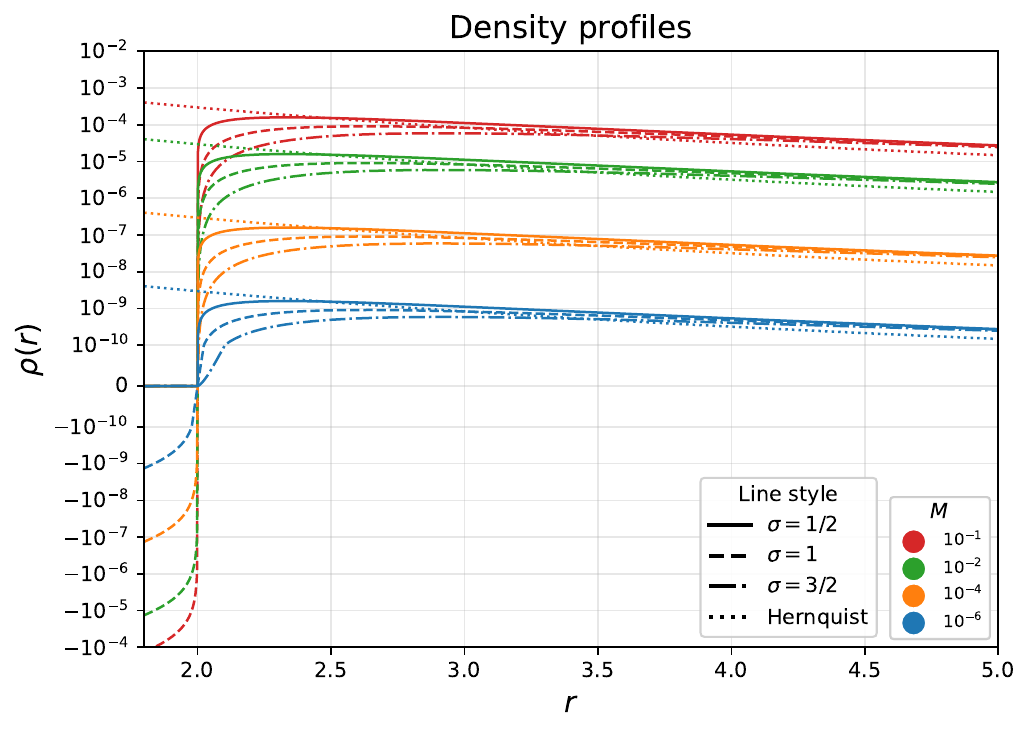}
		\caption{Density profiles $\rho_\sigma(r)$ of Eq.~\eqref{rhosigma} for $\sigma=1/2$ (solid), $1$ (dashed), and $3/2$ (dash-dotted), compared with the pure Hernquist profile \eqref{rhoH} (dotted), for $M=10^{-1},10^{-2},10^{-4},10^{-6}$ (red, green, orange, blue) and $\MBH=a=1$. The vertical axis is symmetric logarithmic. The region $r<2$ lies inside the horizon, where the physical density vanishes; the dashed curves at $r<2$ show the analytic continuation of Eq.~\eqref{rho1}, for illustration only.}\label{fig:profden}
	\end{figure}

	Two remarks are in order. First, the exterior spacetime depends only on $\rho$ for $r>2\MBH$. For $\sigma=1$ the analytic expression \eqref{rho1} becomes negative if it is continued to $r<2\MBH$ (dashed curves in Fig.~\ref{fig:profden}); this continuation is not part of our model, since by (C1) the density vanishes inside the horizon, and it does not affect any of the exterior quantities computed below. Second, for non-integer $\sigma$ the factor $f^\sigma$ is not real for $r<2\MBH$, so that the truncation (C1) is required for the definition of the profile.
	
	Fig.~\ref{fig:profden} illustrates the structure of the family. Since $\rho_\sigma\propto M$, changing $M$ translates the curves vertically on the logarithmic scale, while their shapes depend only on $\sigma$ and $a/\MBH$. Near the horizon the density is suppressed with respect to the Hernquist profile, and the suppression extends to larger radii as $\sigma$ increases. For $a=\MBH$, $\rho_{1/2}$ rises as $\sqrt{r-2\MBH}$ and already equals $\rho_{\rm H}$ at $r\simeq2.5\,\MBH$, whereas $\rho_{3/2}$ is only $\simeq70\%$ of $\rho_{\rm H}$ at $r=3\MBH$. Because the total mass is fixed to $M$, the mass removed near the horizon is redistributed to larger radii: for $r\gg\MBH$, $\rho_\sigma/\rho_{\rm H}\to A_\sigma(\infty)=1+(\sigma+1)\MBH/a$. For $a=\MBH$ this enhancement factor is $2.5$, $3$, and $3.5$ for $\sigma=1/2$, $1$, and $3/2$; at $r=5\MBH$, where $f^\sigma$ is still appreciably smaller than unity, it produces the offset between the colored and dotted curves visible in Fig.~\ref{fig:profden} (e.g., $\rho_1/\rho_{\rm H}=1.8$). For $a\gg\MBH$ one has $A_\sigma\to1$, and the profile reduces to $f^\sigma\rho_{\rm H}$. We note that none of these profiles contains a DM spike \cite{Gondolo:1999ef,Sadeghian:2013laa}: the density near the BH is reduced, not enhanced, with respect to the Hernquist profile. A spike would place more mass at small radii and is expected to enhance the effects discussed below, but it is not considered here.
	
	\subsection{Metric functions}\label{sec:metric}
	
	It is convenient to write $\Phi=\ln f+\PhiH$, where $\PhiH$ is the contribution of the halo, so that
	\begin{equation}
		g_{tt}=-e^{\Phi}=-f(r)\,e^{\PhiH(r)}\equiv-A(r).\label{gttfinal}
	\end{equation}
	Using $1-2m_\sigma/r=f\,(1-D_\sigma)$ with
	\begin{equation}
		D_\sigma(r)\equiv\frac{2Mr\,f(r)^{\sigma}}{(a+r)^2},\label{Dsigma}
	\end{equation}
	Eq.~\eqref{eq:Phi_prime} gives (Appendix~\ref{app:family})
	\begin{align}
		\PhiH'(r)&=\frac{2M\,f(r)^{\sigma-1}}{(a+r)^2\left[1-D_\sigma(r)\right]},\nonumber\\
		\PhiH(r)&=-\int_r^\infty\PhiH'(r')\,dr',\label{PhiHprime}
	\end{align}
	where the integration constant has been fixed by $\PhiH(\infty)=0$, so that $t$ is the proper time of static observers at infinity. The full line element is
	\begin{equation}
		ds^2=-f(r)\,e^{\PhiH(r)}dt^2+\frac{dr^2}{f(r)\left[1-D_\sigma(r)\right]}+r^2d\Omega^2.\label{met1}
	\end{equation}
	For $M\to0$ it reduces to the Schwarzschild metric. At large radii, $\PhiH\simeq-2M/r$ and $m\to\MBH+M$, so that $g_{tt}\simeq-[1-2(\MBH+M)/r]$, consistent with (C2).
	
	Since $(a+r)^2\ge4ar$ and $f<1$, one has $D_\sigma<M/(2a)$ for all $r>2\MBH$. Hence $D_\sigma<1$, and $r=2\MBH$ is the only horizon, whenever $M<2a$ (a sufficient condition). For $\sigma=1$ the necessary and sufficient condition is $M<2a+4\MBH$. All configurations considered in this paper satisfy $M<2a$; for instance, the largest value of $D_\sigma$ in Table~\ref{Tab:ShandPS} is $0.14$. Under this condition $\PhiH'>0$, and therefore
	\begin{equation}
		\PhiH(r)<0\quad\text{and}\quad A(r)<f(r)\quad\text{for all }r\ge2\MBH,\label{PhiHneg}
	\end{equation}
	i.e., the halo increases the gravitational redshift with respect to Schwarzschild. This inequality underlies the results of Sec.~\ref{sec:shadow}.
	
	\textit{Closed form for $\sigma=1$.} In this case $\PhiH'=2M/[r^2+2(a-M)r+a^2+4M\MBH]$, and
	\begin{equation}
		\PhiH^{(\sigma=1)}(r)=-\frac{2M}{\sqrt{\Delta}}\left[\frac{\pi}{2}-\arctan\frac{r+a-M}{\sqrt{\Delta}}\right],\label{PhiH1}
	\end{equation}
	with $\Delta=M(2a+4\MBH-M)>0$, which reproduces the result of Ref.~\cite{Cardoso:2021wlq}. For $\sigma\ne1$ we evaluate Eq.~\eqref{PhiHprime} by numerical quadrature; for $\sigma=1$ the quadrature reproduces Eq.~\eqref{PhiH1} to machine precision.
	
	Both $g_{tt}$ and $g^{rr}$ have simple zeros at $r=2\MBH$, and the surface gravity associated with the Killing vector $\partial_t$ is
	\begin{equation}
		\kappa_s=\frac{e^{\PhiH(2\MBH)/2}}{4\MBH}<\frac{1}{4\MBH}.\label{kappas}
	\end{equation}
	For $a=\MBH$ and $M=0.1\,\MBH$, the reduction factor $e^{\PhiH(2\MBH)/2}$ is $0.942$, $0.967$, and $0.976$ for $\sigma=1/2$, $1$, and $3/2$.
	
	\subsection{Horizon regularity}\label{sec:regularity}
	
	Near the horizon, $r=2\MBH+\epsilon$ with $\epsilon\to0^+$, one has $\rho\propto\epsilon^\sigma$, $m-\MBH=O(\epsilon^{\sigma+1})$, $m'\propto\epsilon^\sigma$, and $r-2m\simeq\epsilon$. Equation~\eqref{eq:pt} then gives $p_t\propto\epsilon^{\sigma-1}$, and Eq.~\eqref{Ricci} yields
	\begin{equation}
		R\simeq-\frac{(\sigma+1)\,M}{(2\MBH)^{\sigma}(a+2\MBH)^2}\,\epsilon^{\sigma-1},\label{Rnear}
	\end{equation}
	that is,
	\begin{align}
		R_{\sigma=1/2}&\simeq-\frac{3\sqrt2\,M}{4\sqrt{\MBH}\,(a+2\MBH)^2}\,\epsilon^{-1/2},\nonumber\\
		R_{\sigma=1}&\to-\frac{M}{\MBH(a+2\MBH)^2},\nonumber\\
		R_{\sigma=3/2}&\simeq-\frac{5\sqrt2\,M}{8\MBH^{3/2}(a+2\MBH)^2}\,\epsilon^{1/2}.\nonumber
	\end{align}
	In the Kretschmann scalar \eqref{Kretsch}, the Schwarzschild term $48m^2/r^6$ is finite, while the matter terms scale as $m'/\epsilon\propto\epsilon^{\sigma-1}$ and $m'^2/\epsilon^2\propto\epsilon^{2\sigma-2}$. Hence $K$ diverges at the horizon for $\sigma<1$ and is finite for $\sigma\ge1$.
	
	The same criterion follows from the energy density measured by a radially infalling observer with conserved energy $E=A\,\dot t$. Since $p_r=0$,
	\begin{equation}
		T_{\mu\nu}u^\mu u^\nu=\frac{E^2\rho}{A}\propto\epsilon^{\sigma-1}\qquad(\epsilon\to0^+),\label{rhoff}
	\end{equation}
	which diverges at the horizon for $\sigma<1$ and remains finite for $\sigma\ge1$. We therefore conclude that:
	(i) for $\sigma<1$, including $\sigma=1/2$, the horizon carries a curvature singularity, at which freely falling observers measure an infinite energy density;
	(ii) for $\sigma=1$, curvature invariants and $p_t$ are finite at the horizon but jump to their vacuum values across it, so that the metric is not $C^2$ there;
	(iii) for $\sigma>1$, they approach their vacuum values continuously.
	The range $\sigma\ge1$ is thus the physically admissible one. We nevertheless retain $\sigma=1/2$ as a reference case that brackets the dependence on the near-horizon behavior, bearing in mind that it is singular at the horizon. As discussed in Sec.~\ref{sec:EC}, the shadow and the QNMs probe the region $r\gtrsim3\MBH$, and the singular layer affects them only through the small mass it contains.
	
	\subsection{Physical interpretation of the closure}\label{sec:validity}
	
	The Einstein-cluster interpretation imposes a further restriction. For a swarm of particles moving on circular geodesics with local speed $v$ measured by static observers, averaging over the orbital orientations gives $p_t=\rho v^2/2$, while circular geodesic motion requires $v^2=r\Phi'/2$. Combining these relations with Eq.~\eqref{eq:Phi_prime} reproduces Eq.~\eqref{eq:pt} with
	\begin{equation}
		v^2(r)=\frac{m(r)}{r-2m(r)},\qquad v<1\iff r>3m(r).\label{vsq}
	\end{equation}
	The constituents are therefore subluminal only outside the circular photon orbit, $r>\rPS$ with $\rPS=3m(\rPS)$ [Eq.~\eqref{rPS3m} below]. In the layer $2\MBH<r<\rPS$, the stress tensor cannot be realized by particles on circular geodesics and must be regarded as an effective anisotropic fluid. Moreover, as in Schwarzschild, circular timelike geodesics between the photon sphere and the innermost stable circular orbit (located near $6\MBH$ for small $M/a$) are unstable. These features are shared by all Einstein-cluster models in which the matter extends down to the horizon, including the $\sigma=1$ solution of Ref.~\cite{Cardoso:2021wlq} (see also Refs.~\cite{Datta:2023zmd,Maeda:2024tsg}). The photon-sphere-based observables considered in this paper, however, are dominated by the geometry at $r\gtrsim\rPS$, as we quantify below.
	
	\section{Energy conditions}\label{sec:EC}
	
	For $T^\mu{}_\nu=\mathrm{diag}(-\rho,p_r,p_t,p_t)$, the null (NEC), weak (WEC), strong (SEC), and dominant (DEC) energy conditions read \cite{Guo:2022ghl}
	\begin{itemize}
		\item[(i)] NEC: $\rho+p_i\ge0$ $(i=r,t)$;
		\item[(ii)] WEC: $\rho\ge0$ and the NEC;
		\item[(iii)] SEC: $\rho+p_r+2p_t\ge0$ and the NEC;
		\item[(iv)] DEC: $\rho\ge|p_i|$, i.e., $\rho\ge0$ and $\rho\pm p_i\ge0$.
	\end{itemize}
	Following the notation of Ref.~\cite{Guo:2022ghl}, $\mathrm{WEC}_{1,2}=\mathrm{NEC}_{1,2}$ denote $\rho+p_i\ge0$ for $i=r,t$, $\mathrm{WEC}_3=\mathrm{DEC}_1$ denotes $\rho\ge0$, and $\mathrm{DEC}_{2,3}$ denote $\rho\pm p_i\ge0$ for $i=r,t$.
	
	Inside the horizon the spacetime is vacuum, and all conditions hold trivially. Outside, $\rho\ge0$ by construction, and Eq.~\eqref{eq:pt} with $m>0$ and $r>2m$ gives $p_t\ge0$. Together with $p_r=0$, this implies that for $i=r$, the WEC, NEC, and SEC are satisfied everywhere, since all the relevant combinations are sums of non-negative terms. The nontrivial combinations are
	\begin{align}
		\mathrm{WEC}_2=\mathrm{NEC}_2&:\quad\rho+p_t=\rho\,\frac{2r-3m}{2(r-2m)},\label{WEC2}\\
		\mathrm{SEC}&:\quad\rho+p_r+2p_t=\rho\,\frac{r-m}{r-2m},\label{SEC}\\
		\mathrm{DEC}_3&:\quad\rho-p_t=\rho\,\frac{2r-5m}{2(r-2m)},\label{DEC3}
	\end{align}
	while $\mathrm{DEC}_2$, $\rho\pm p_r=\rho\ge0$, is trivially satisfied. Equation~\eqref{DEC3} shows that the DEC is violated if and only if $2r<5m(r)$. Since $m\ge\MBH$, this happens in a layer $2\MBH<r<r_{\rm DEC}$, where $r_{\rm DEC}=5m(r_{\rm DEC})/2>5\MBH/2$. The values of $r_{\rm DEC}$ are listed in Table~\ref{Tab:ShandPS}: for $a=\MBH$ and $0.01\le M/a\le0.6$ they lie in the range $2.50\lesssim r_{\rm DEC}/\MBH\lesssim2.58$, increasing with $M$ and decreasing with $\sigma$.
	
	\begin{figure}[t]
		\includegraphics[width=0.95\columnwidth]{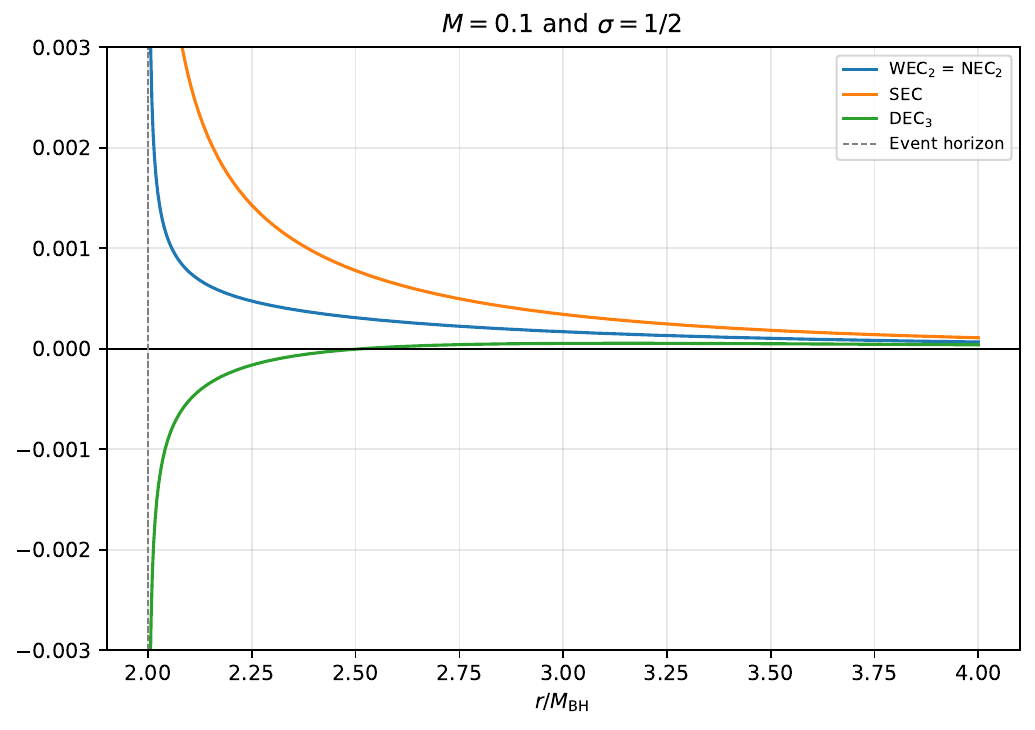}\\
		\includegraphics[width=0.95\columnwidth]{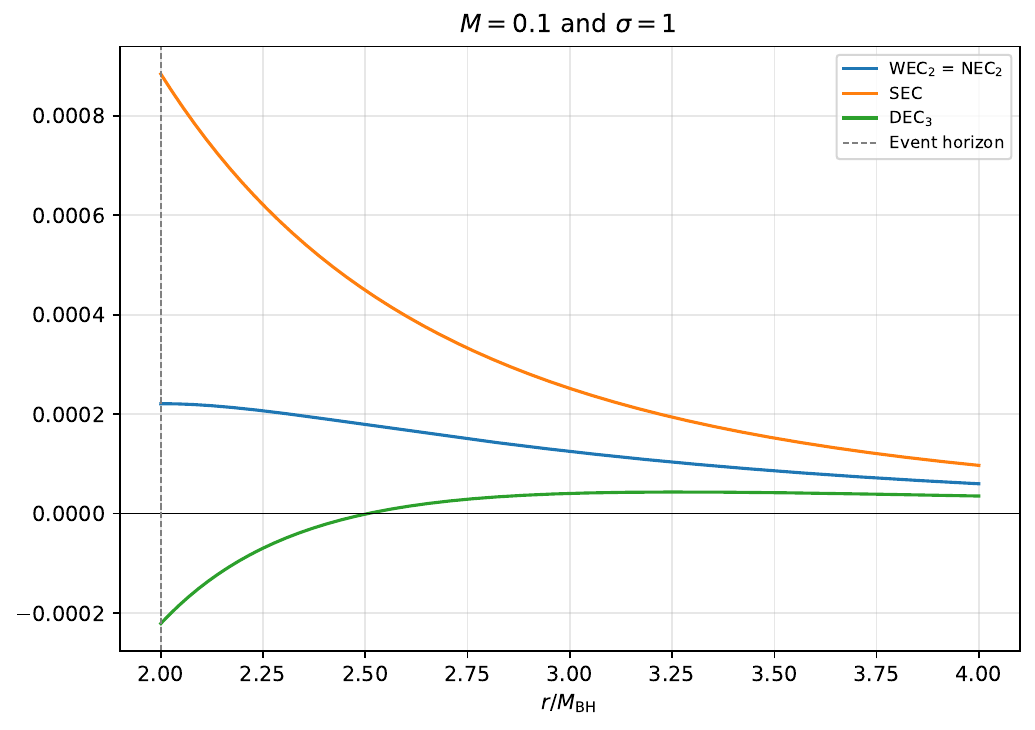}\\
		\includegraphics[width=0.95\columnwidth]{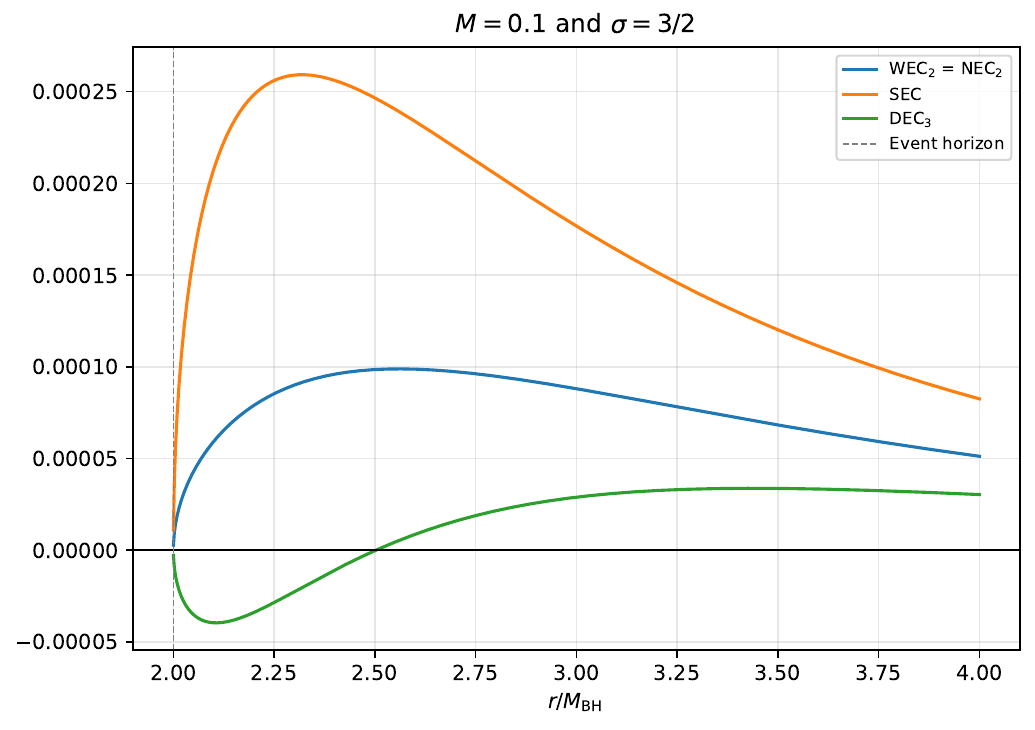}
		\caption{Energy-condition combinations $\rho+p_t$ ($\mathrm{WEC}_2=\mathrm{NEC}_2$), $\rho+p_r+2p_t$ (SEC), and $\rho-p_t$ ($\mathrm{DEC}_3$), Eqs.~\eqref{WEC2}-\eqref{DEC3}, as functions of $r/\MBH$ for $M=0.1\,\MBH$, $a=\MBH$, and (a) $\sigma=1/2$, (b) $\sigma=1$, (c) $\sigma=3/2$. The vertical dashed line marks the horizon. In panel (a) the vertical axis is truncated, since all three combinations diverge at the horizon.}\label{fig:EC01}
	\end{figure}
	
	Fig.~\ref{fig:EC01} displays the combinations \eqref{WEC2}-\eqref{DEC3} for $M=0.1\,\MBH$ and $a=\MBH$. For all three values of $\sigma$, $\mathrm{DEC}_3$ is the only condition that is violated, and the sign change occurs at $r_{\rm DEC}\simeq2.51\,\MBH$, $2.51\,\MBH$, and $2.50\,\MBH$ for $\sigma=1/2$, $1$, and $3/2$. Just outside the horizon the ordering is $\mathrm{SEC}>\mathrm{WEC}_2>0>\mathrm{DEC}_3$, while at large $r$ all combinations are positive and approach $\rho$, since $p_t/\rho\to0$. The near-horizon behavior follows from $p_t\propto\epsilon^{\sigma-1}$ and $\rho\propto\epsilon^\sigma$. For $\sigma=1/2$, all three combinations diverge as $r\to2\MBH^+$, the first two to $+\infty$ and $\mathrm{DEC}_3$ to $-\infty$, consistently with the curvature singularity of Sec.~\ref{sec:regularity}. For $\sigma=1$, $p_t$ tends to a finite value while $\rho\to0$, so that $\mathrm{SEC}=2\,\mathrm{WEC}_2=-2\,\mathrm{DEC}_3=2p_t$ at the horizon, and these values jump to zero across it. For $\sigma=3/2$, all combinations vanish at the horizon, and their extrema move outward, reflecting the broader suppression region of the density seen in Fig.~\ref{fig:profden}.
	
	The DEC violation is the energy-condition counterpart of the superluminal velocities discussed in Sec.~\ref{sec:validity}. Using Eq.~\eqref{vsq}, $\rho\ge p_t$ is equivalent to $v^2\le2$, which is a weaker requirement than $v<1$; the DEC violation is therefore confined to a region, $r<r_{\rm DEC}\simeq2.5\,\MBH$, that lies entirely inside the region $r<\rPS\simeq3\MBH$ in which the particle interpretation already fails. The photon sphere ($\rPS\simeq3.00$-$3.25\,\MBH$), the shadow, and the peak of the QNM potentials ($r_0\simeq3.0$-$3.3\,\MBH$) all lie outside this region. The inner layer influences them only through the halo mass it contains, which for $a=\MBH$ amounts to a fraction $3^{1-\sigma}\MBH^2/(a+3\MBH)^2$ of $M$ inside $r=3\MBH$, i.e., $11\%$, $6\%$, and $4\%$ for $\sigma=1/2$, $1$, and $3/2$ (Sec.~\ref{sec:PS}).
	
	\section{Photon sphere and shadow}\label{sec:shadow}
	
	\subsection{Null geodesics and photon sphere}\label{sec:PS}
	
	Owing to spherical symmetry we restrict to the equatorial plane, $\theta=\pi/2$. The conserved energy and angular momentum of a photon are
	\begin{equation}
		E=A(r)\,\dot t,\qquad L=r^2\dot\phi,
	\end{equation}
	where a dot denotes the derivative with respect to an affine parameter and $A=fe^{\PhiH}$. The null condition gives
	\begin{equation}
		\dot r^2=\frac{1-2m/r}{A(r)}\left[E^2-V_{\rm eff}(r)\right],\qquad V_{\rm eff}=\frac{A(r)\,L^2}{r^2}.\label{Veff}
	\end{equation}
	In terms of the impact parameter $b=L/E$ and $u=1/r$, the orbit equation used for ray tracing reads
	\begin{equation}
		\left(\frac{du}{d\phi}\right)^2=\frac{1-2m/r}{A(r)}\left[\frac{1}{b^2}-A(r)\,u^2\right].\label{orbit}
	\end{equation}
	Circular photon orbits satisfy $V_{\rm eff}=E^2$ and $V_{\rm eff}'=0$, i.e., $rA'=2A$, or
	\begin{equation}
		r\,\Phi'(r)=2.\label{PScond}
	\end{equation}
	With the closure \eqref{eq:Phi_prime}, this condition becomes
	\begin{equation}
		\rPS=3\,m(\rPS).\label{rPS3m}
	\end{equation}
	This relation holds for any mass function within the Einstein-cluster closure, not only for the family \eqref{msigma}. Since $m(r)>\MBH$ outside the horizon for $M>0$,
	\begin{equation}
		\rPS>3\MBH\qquad(M>0).\label{rPSbound}
	\end{equation}
	For all the configurations considered here, $r-3m(r)$ has a single root for $r>2\MBH$, so $V_{\rm eff}$ has a single maximum. To first order in $M$, $\delta\rPS=3\,\delta m(3\MBH)$, because the correction $3m'\,\delta\rPS$ is of order $M^2$. Using $f(3\MBH)=1/3$ in Eq.~\eqref{msigma} gives
	\begin{equation}
		\delta\rPS=3^{\,2-\sigma}\,\frac{M\,\MBH^2}{(a+3\MBH)^2},\label{drPS}
	\end{equation}
	i.e., $3\sqrt3\,M\MBH^2/(a+3\MBH)^2$, $3M\MBH^2/(a+3\MBH)^2$, and $\sqrt3\,M\MBH^2/(a+3\MBH)^2$ for $\sigma=1/2$, $1$, and $3/2$. Equivalently, the photon sphere moves out by three times the halo mass enclosed at $r=3\MBH$, which is the fraction $3^{1-\sigma}\MBH^2/(a+3\MBH)^2$ of $M$ quoted in Sec.~\ref{sec:EC}.
	
	\subsection{Shadow}\label{sec:sh}
	
\begin{figure}[t]
		\includegraphics[width=0.45\textwidth]{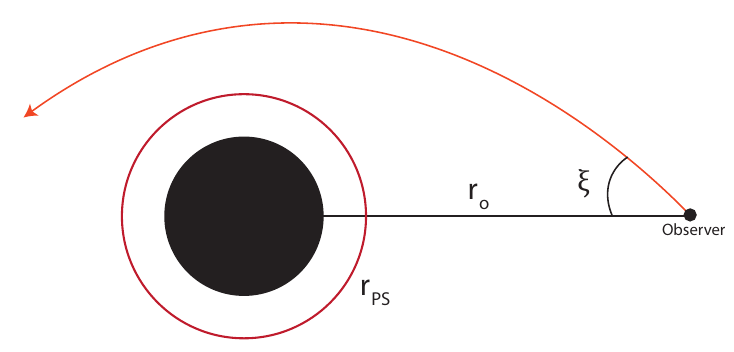}
		\caption{Schematic of the shadow construction: an observer at $r_o$ sees the edge of the shadow at the angular radius $\xi$, set by the ray that asymptotically approaches the photon sphere of radius $\rPS$ \cite{Perlick:2021aok}.}\label{fig:shschm}
	\end{figure}

	A static observer at radius $r_o$ sees the edge of the shadow, formed by rays that asymptotically approach the photon sphere, at the angular radius $\xi$ given by \cite{Synge:1966okc,Perlick:2021aok} (Fig.~\ref{fig:shschm})
	\begin{align}
		\sin\xi&=\frac{b_c\sqrt{A(r_o)}}{r_o},\nonumber\\
		b_c&=\frac{\rPS}{\sqrt{A(\rPS)}}=\frac{\rPS\,e^{-\PhiH(\rPS)/2}}{\sqrt{f(\rPS)}}.\label{sinxi}
	\end{align}
	Equivalently, $b_c^{-2}=\max_{r>2\MBH}\,[A(r)/r^2]$: photons from infinity with $b>b_c$ are deflected, and those with $b<b_c$ are captured. For an observer at a distance $D\gg a$, the angular radius of the shadow is $b_c/D$, and we define the shadow radius as $\rSh\equiv b_c$.
	
	Because $A(r)<f(r)$ for every $r>2\MBH$ [Eq.~\eqref{PhiHneg}],
	\begin{equation}
		b_c^2=\min_{r>2\MBH}\frac{r^2}{A(r)}>\min_{r>2\MBH}\frac{r^2}{f(r)}=27\MBH^2.\label{bcbound}
	\end{equation}
	Hence $\rSh>3\sqrt3\,\MBH$ for every $M>0$ and every $\sigma$, provided $D_\sigma<1$. Moreover, the integrand $\PhiH'$ in Eq.~\eqref{PhiHprime} is, at each $r$, a strictly increasing function of $M$ and a strictly decreasing function of $a$ and of $\sigma$ (recall that $f<1$, and that $D_\sigma$ increases with $M$ and decreases with $a$ and $\sigma$). Since $b_c^{-2}$ is the maximum of $A/r^2=f e^{\PhiH}/r^2$, it follows that, at fixed $\MBH$,
	\begin{equation}
		\frac{\partial b_c}{\partial M}>0,\qquad\frac{\partial b_c}{\partial a}<0,\qquad\frac{\partial b_c}{\partial \sigma}<0.\label{monotonic}
	\end{equation}
	These statements rely only on Eqs.~\eqref{eq:Phi_prime}, \eqref{msigma}, and \eqref{PhiHprime}, and they hold at fixed $\MBH$. Physically, the shadow is enlarged both by the outward shift of the photon sphere and, more importantly, by the additional redshift $e^{\PhiH}<1$ accumulated by the halo outside the photon sphere.
	
	To first order in $M$, the shift of $\rPS$ does not contribute to $b_c$, because $r^2/A$ is stationary at $\rPS$. To this order, $\PhiH'\simeq2\,\delta m(r)/[r^2f(r)^2]$ with $\delta m=m-\MBH$, and Eq.~\eqref{sinxi} gives
	\begin{align}
		\frac{\delta\rSh}{3\sqrt3\,\MBH}&\simeq-\frac12\PhiH(3\MBH)\Big|_{O(M)}=\int_{3\MBH}^{\infty}\frac{\delta m(r)}{r^2f(r)^2}\,dr\nonumber\\
		&=M\int_{3\MBH}^\infty\frac{f(r)^{\sigma-1}}{(a+r)^2}\,dr\equiv M\,I_\sigma(a).\label{firstorder}
	\end{align}
	For $\sigma=1$, $I_1=1/(a+3\MBH)$. For $a=\MBH$ one finds $I_\sigma\MBH=0.309$, $0.250$, and $0.207$ for $\sigma=1/2$, $1$, and $3/2$. For $a\gg\MBH$, $I_\sigma\simeq(a+3\MBH)^{-1}[1-2(\sigma-1)(\MBH/a)\ln(a/3\MBH)]$, so that the dependence on $\sigma$ disappears as $(\MBH/a)\ln(a/\MBH)$: the shift is then controlled by $M/a$ alone, in agreement with the leading-order behavior $b_c\simeq3\sqrt3\MBH(1+M/a)$ of the $\sigma=1$ solution in the regime $a\gg\MBH$ (see, e.g., Ref.~\cite{Cardoso:2021wlq}). Equation~\eqref{firstorder} makes explicit that the relevant parameter is $M/(a+3\MBH)$ rather than $M/a$, and that the shadow receives contributions from the whole halo outside $r=3\MBH$, weighted by $1/(r^2f^2)$. The first-order formula reproduces the full numerical shift to within $0.3\%$ of its value for $M=0.01\,\MBH$ and within $2$-$3\%$ for $M=0.1\,\MBH$ (Appendix~\ref{app:numerics}).
	
	\subsection{Numerical results}\label{sec:shres}
	
\begin{figure*}[t]
		\centering
		\includegraphics[width=0.48\textwidth]{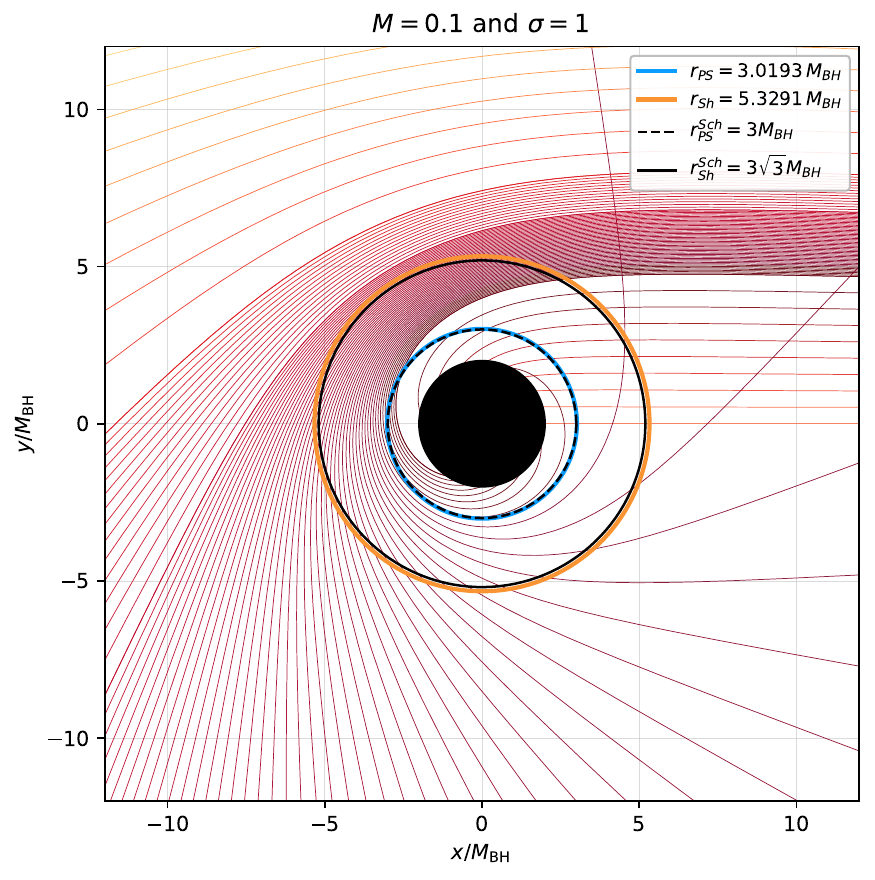}
		\hfill
		\includegraphics[width=0.48\textwidth]{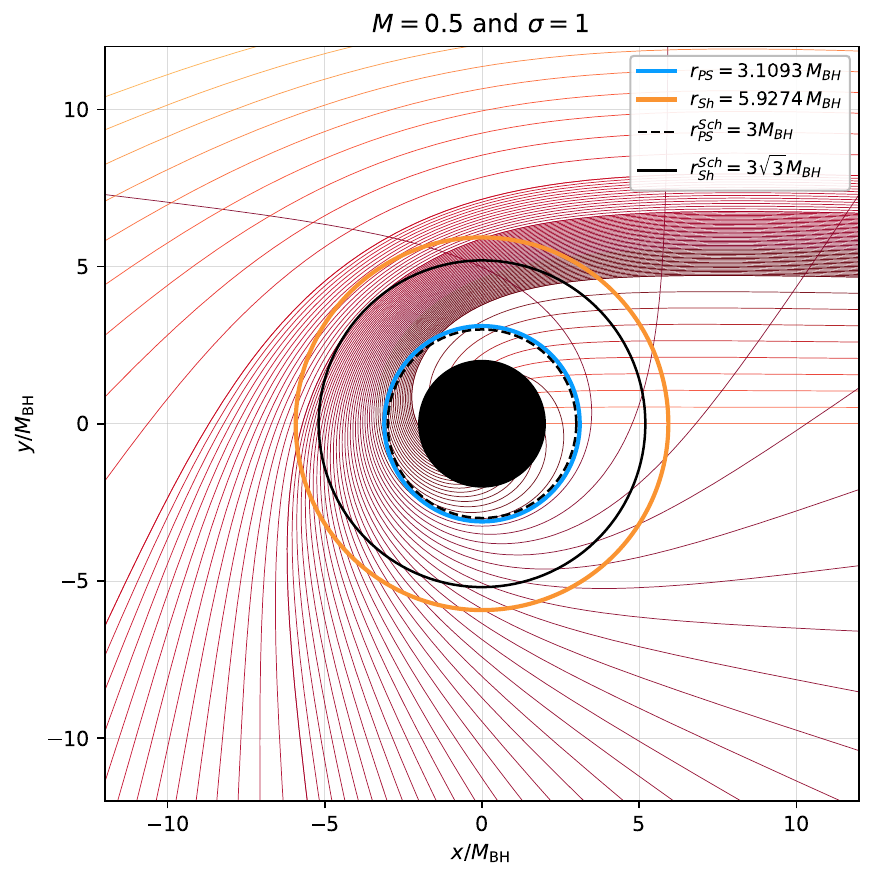}\\[2mm]
		\includegraphics[width=0.48\textwidth]{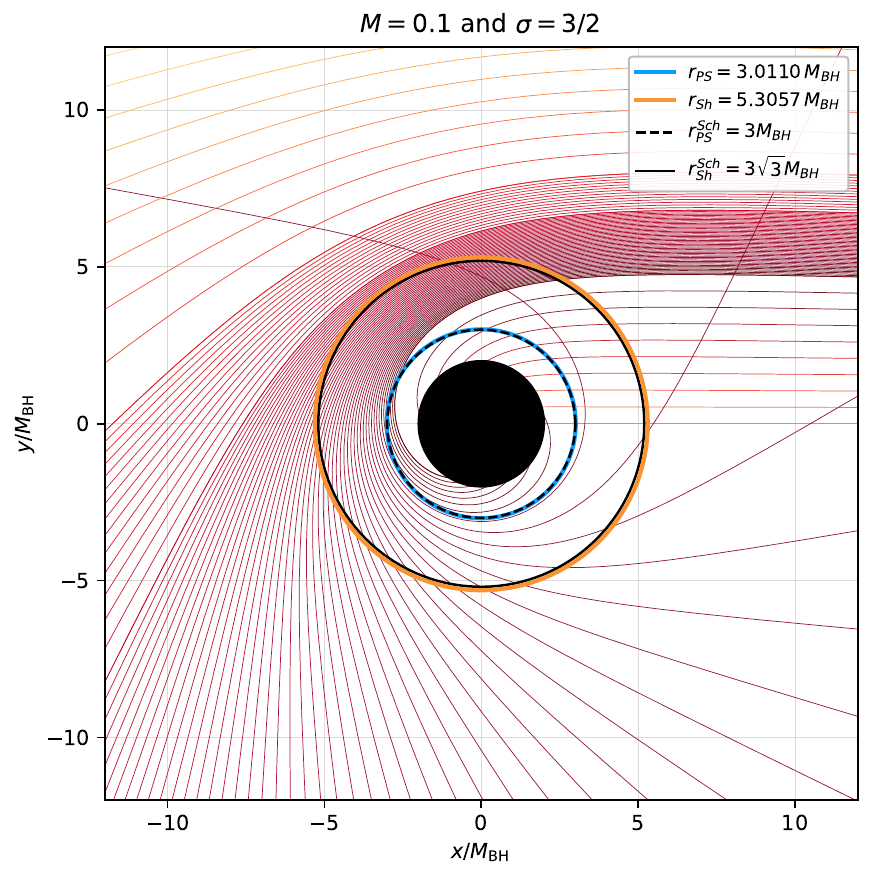}
		\hfill
		\includegraphics[width=0.48\textwidth]{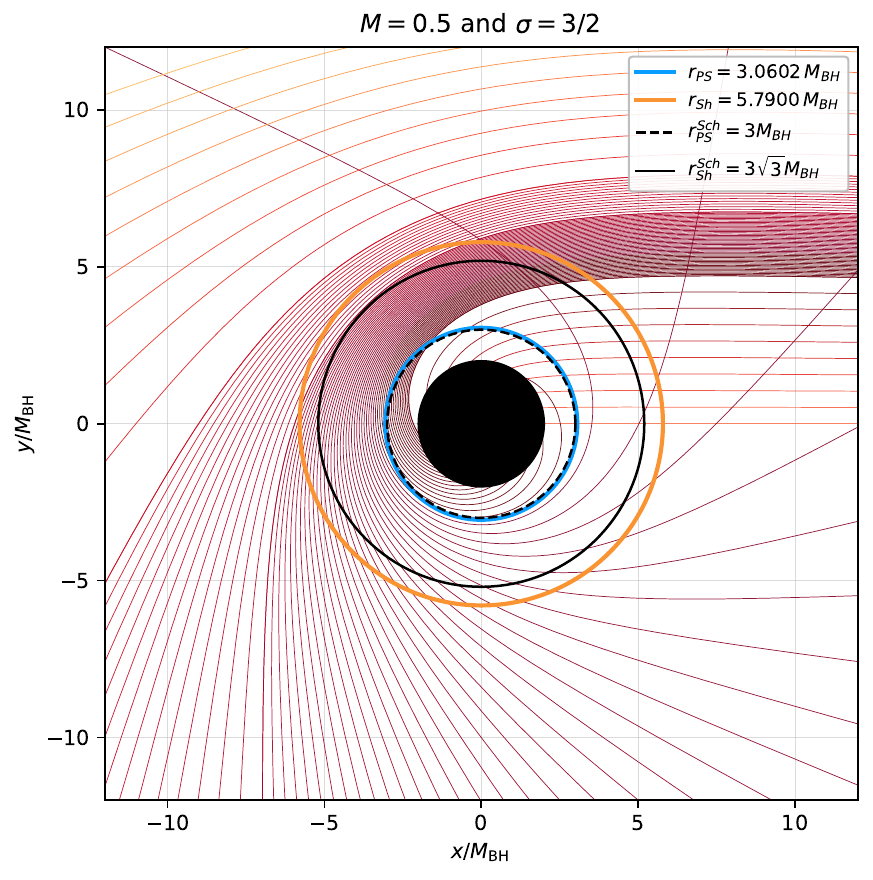}
		\caption{Null geodesics (thin lines) of the metric \eqref{met1} obtained from Eq.~\eqref{orbit} for a parallel beam incident from the right, for $\MBH=a=1$, $M=0.1$ (left) and $M=0.5$ (right), and $\sigma=1$ (top) and $\sigma=3/2$ (bottom). The blue and orange circles have radii $\rPS$ and $b_c=\rSh$, respectively; the dashed and solid black circles show the corresponding Schwarzschild values $3\MBH$ and $3\sqrt3\,\MBH$. The black disk is the BH interior, $r\le2\MBH$. Rays with $b<b_c$ are captured, whereas rays with $b$ slightly above $b_c$ wind around the photon sphere before escaping.}
		\label{fig:Shad}
	\end{figure*}

	The photon-sphere and shadow radii for an observer at infinity are listed in Table~\ref{Tab:ShandPS} and visualized, together with ray tracing based on Eq.~\eqref{orbit}, in Fig.~\ref{fig:Shad}.
	
	\begin{table}[t]
		\caption{Photon-sphere radius $\rPS$, shadow radius $\rSh=b_c$, relative deviation $\delta_{\rm Sh}\equiv\rSh/(3\sqrt3\,\MBH)-1$ from the Schwarzschild value, shadow radius in units of $\MADM=\MBH+M$, and outer radius $r_{\rm DEC}$ of the layer in which the DEC is violated (Sec.~\ref{sec:EC}), for $a=\MBH$. The quantities $\rPS$, $\rSh$, and $r_{\rm DEC}$ are given in units of $\MBH$.}
		\label{Tab:ShandPS}
		\begin{ruledtabular}
			\begin{tabular}{ccccccc}
				$\sigma$ & $M/a$ & $\rPS$ & $\rSh$ & $\delta_{\rm Sh}$ (\%) & $\rSh/\MADM$ & $r_{\rm DEC}$\\
				\colrule
				1/2 & 0.01 & 3.00326 & 5.21225 & $+0.31$ & 5.16065 & 2.501\\
				& 0.1  & 3.03375 & 5.36111 & $+3.17$ & 4.87373 & 2.512\\
				& 0.5  & 3.19995 & 6.11828 & $+17.75$ & 4.07886 & 2.567\\
				& 0.6  & 3.25138 & 6.33511 & $+21.92$ & 3.95945 & 2.584\\
				\colrule
				1   & 0.01 & 3.00188 & 5.20917 & $+0.25$ & 5.15760 & 2.501\\
				& 0.1  & 3.01929 & 5.32906 & $+2.56$ & 4.84460 & 2.505\\
				& 0.5  & 3.10931 & 5.92742 & $+14.07$ & 3.95162 & 2.528\\
				& 0.6  & 3.13575 & 6.09592 & $+17.32$ & 3.80995 & 2.534\\
				\colrule
				3/2 & 0.01 & 3.00108 & 5.20692 & $+0.21$ & 5.15537 & 2.500\\
				& 0.1  & 3.01105 & 5.30572 & $+2.11$ & 4.82338 & 2.502\\
				& 0.5  & 3.06020 & 5.79000 & $+11.43$ & 3.86000 & 2.512\\
				& 0.6  & 3.07393 & 5.92393 & $+14.01$ & 3.70246 & 2.515\\
			\end{tabular}
		\end{ruledtabular}
	\end{table}
	
	Table~\ref{Tab:ShandPS} and Fig.~\ref{fig:Shad} confirm the analytical results of Sec.~\ref{sec:sh}. The photon sphere moves outward only slightly ($\rPS\lesssim3.25\,\MBH$ even for $M/a=0.6$), as dictated by Eqs.~\eqref{rPS3m} and \eqref{drPS}, because only the fraction of the halo mass enclosed within $\rPS$ contributes. The shadow responds more strongly, because it is also affected by the redshift $e^{\PhiH}$, which receives contributions from the entire halo outside the photon sphere [Eq.~\eqref{firstorder}]. For $M/a=0.1$, the relative shift of the shadow ($2.1$-$3.2\%$) is three to six times larger than that of the photon sphere ($0.37$-$1.1\%$). In the ray-tracing panels, the numerical circles are almost indistinguishable from the Schwarzschild ones for $M=0.1$, whereas for $M=0.5$ the shadow edge is visibly displaced, by $11\%$ for $\sigma=3/2$ and $14\%$ for $\sigma=1$, while the photon sphere moves by only $2$-$4\%$. The ordering of the shifts, $\sigma=1/2>1>3/2$ at fixed $M/a$, is the one required by Eq.~\eqref{monotonic}. It reflects the fact that the halo mass enclosed within a given radius, $\delta m(r)=Mr^2f^{\sigma+1}/(a+r)^2$, is larger for smaller $\sigma$ at every $r$, and that Eq.~\eqref{firstorder} weights this mass most heavily at small radii.
	
	\subsection{Comparison with EHT observations}\label{sec:EHT}
	
	For Sgr\,A$^*$, combining the EHT image \cite{EventHorizonTelescope:2022wkp,EventHorizonTelescope:2022xqj} with the mass-to-distance ratio inferred from the Keck and VLTI monitoring of stellar orbits yields $4.55\lesssim\rSh/M_{\rm obs}\lesssim5.22$ at $1\sigma$ and $4.21\lesssim\rSh/M_{\rm obs}\lesssim5.56$ at $2\sigma$ \cite{Vagnozzi:2022moj}, where $M_{\rm obs}$ denotes the mass used as a prior. For M87$^*$, the shadow diameter $d_{\rm sh}=(11.0\pm1.5)\,M_{\rm obs}$ \cite{EventHorizonTelescope:2019dse,EventHorizonTelescope:2019ggy,Bambi:2019tjh} gives $4.75\lesssim\rSh/M_{\rm obs}\lesssim6.25$ at $1\sigma$. In what follows we adopt the $2\sigma$ interval for Sgr\,A$^*$ and the $1\sigma$ interval for M87$^*$; the different confidence levels should be kept in mind when comparing the two sources.
	
\begin{figure}[t]
		\centering
		\includegraphics[width=0.45\textwidth]{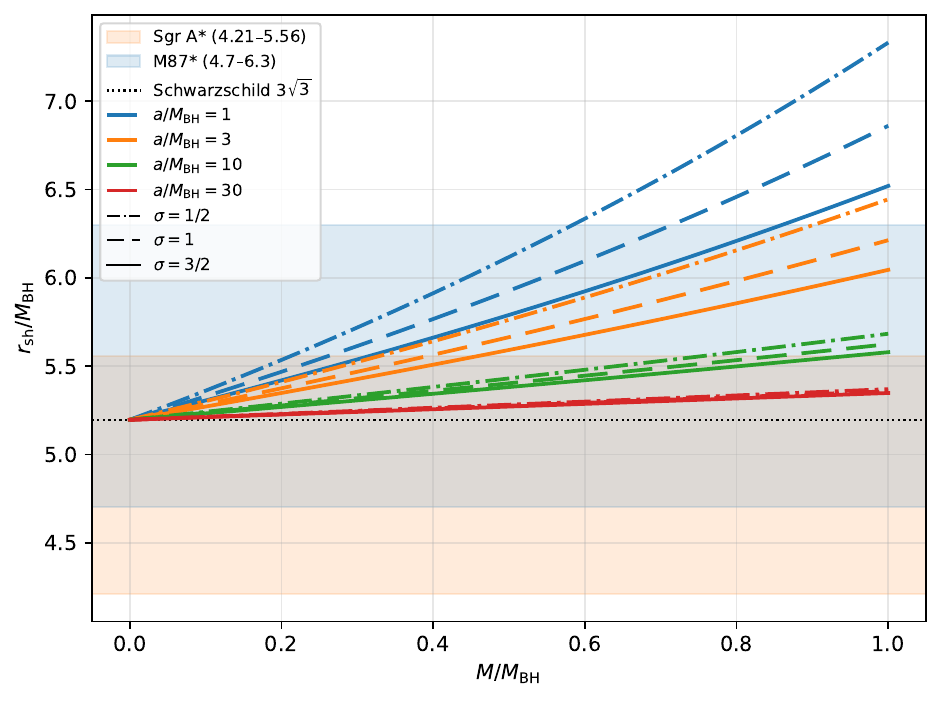}
		\caption{Shadow radius as a function of the halo mass for $a/\MBH=1$, $3$, $10$, and $30$ (blue, orange, green, red), and $\sigma=1/2$ (dash-dotted), $1$ (dashed), and $3/2$ (solid). (a) $\rSh/\MBH$, convention (i); the dotted line is the Schwarzschild value $3\sqrt3$. (b) $\rSh/\MADM$, convention (ii); the dotted line, $3\sqrt3\,\MBH/\MADM$, shows the effect of the change of mass unit alone, without any halo-induced lensing. The shaded bands are the Sgr\,A$^*$ ($2\sigma$) and M87$^*$ ($1\sigma$) intervals.}\label{fig:Band}
	\end{figure}

	The bounds depend on which mass of the model is identified with $M_{\rm obs}$. We consider two conventions, whose results are summarized in Table~\ref{Tab:bounds} and Fig.~\ref{fig:Band}:
	\begin{itemize}
		\item[(i)] $M_{\rm obs}=\MBH$. The model prediction is $\rSh/\MBH$, which increases with $M$; the constraint comes from the upper edges of the observed intervals, and Sgr\,A$^*$ is the more constraining source. For general $a$ and $\sigma=1$, Eq.~\eqref{firstorder} translates the Sgr\,A$^*$ condition $\delta\rSh/\rSh^{\rm Schw}<0.070$ into $M/(a+3\MBH)\lesssim0.07$.
		\item[(ii)] $M_{\rm obs}=\MADM$. The model prediction is $\rSh/\MADM$, which decreases with $M$; the constraint comes from the lower edges, and M87$^*$ is the more constraining source.
	\end{itemize}
	
	For the compact halos considered here, convention (ii) is the one consistent with the way the mass priors are obtained. For Sgr\,A$^*$, the mass is determined from the orbits of the S stars, whose pericenters lie at $\gtrsim10^3$ gravitational radii (about $3\times10^3\,\MBH$ for S2) \cite{GRAVITY:2020gka}; for M87$^*$, the EHT mass is consistent with stellar-dynamical estimates obtained on much larger scales \cite{EventHorizonTelescope:2019ggy}. For $a\lesssim30\,\MBH$, at least $98\%$ of the halo mass lies within $3\times10^3\,\MBH$, so that these measurements determine $\MADM$ rather than $\MBH$. Convention (i) would apply if $\MBH$ were measured independently at radii $r\lesssim a$, which is not possible at present; we nevertheless report it because it isolates the effect of the halo on light propagation at fixed BH mass, and because it is the convention implicitly adopted when shadow radii are quoted in units of the BH mass. For halos with $a$ comparable to the S-star orbits, the dynamical mass includes only part of the halo, and neither convention applies; such halos are instead constrained by the absence of an extended mass in the orbit of S2 \cite{GRAVITY:2020gka}.
	
	\begin{table*}[t]
		\caption{Upper bounds on $M/\MBH$ from the EHT shadow sizes, for $\sigma=1/2$, $1$, and $3/2$ and several values of $a/\MBH$. In convention (i) the mass prior is identified with $\MBH$ and $\rSh/\MBH$ must lie below the upper edge of the observed interval; in convention (ii) it is identified with $\MADM=\MBH+M$ and $\rSh/\MADM$ must lie above the lower edge. We use the $2\sigma$ interval for Sgr\,A$^*$ and the $1\sigma$ interval for M87$^*$. An entry ``$>1$'' means that no bound is obtained for $M\le\MBH$. For $a\to\infty$, convention (ii) reduces to the requirement $3\sqrt3\,\MBH/\MADM$ within the observed interval, independently of $\sigma$; this limit is shown for reference only, since for very extended halos the dynamical mass does not include the whole halo (see text).}
		\label{Tab:bounds}
		\begin{ruledtabular}
			\begin{tabular}{ccccccccccccc}
				& \multicolumn{6}{c}{(i) $M_{\rm obs}=\MBH$: $\rSh/\MBH<$ upper edge} & \multicolumn{6}{c}{(ii) $M_{\rm obs}=\MADM$: $\rSh/\MADM>$ lower edge}\\
				& \multicolumn{3}{c}{Sgr\,A$^*$ ($<5.56$)} & \multicolumn{3}{c}{M87$^*$ ($<6.25$)} & \multicolumn{3}{c}{Sgr\,A$^*$ ($>4.21$)} & \multicolumn{3}{c}{M87$^*$ ($>4.75$)}\\
				$a/\MBH$ & $\sigma=\frac12$ & $1$ & $\frac32$ & $\frac12$ & $1$ & $\frac32$ & $\frac12$ & $1$ & $\frac32$ & $\frac12$ & $1$ & $\frac32$\\
				\colrule
				1        & 0.21 & 0.26 & 0.32 & 0.56 & 0.69 & 0.83 & 0.41 & 0.35 & 0.32 & 0.14 & 0.13 & 0.12\\
				3        & 0.33 & 0.40 & 0.46 & 0.87 & $>1$ & $>1$ & 0.32 & 0.30 & 0.29 & 0.12 & 0.12 & 0.11\\
				10       & 0.76 & 0.86 & 0.95 & $>1$ & $>1$ & $>1$ & 0.26 & 0.26 & 0.26 & 0.10 & 0.10 & 0.10\\
				30       & $>1$ & $>1$ & $>1$ & $>1$ & $>1$ & $>1$ & 0.24 & 0.24 & 0.24 & 0.097 & 0.097 & 0.097\\
				$\infty$ & \multicolumn{6}{c}{no bound} & 0.23 & 0.23 & 0.23 & 0.094 & 0.094 & 0.094\\
			\end{tabular}
		\end{ruledtabular}
	\end{table*}
	
	Fig.~\ref{fig:Band}(a) shows the dependence of convention (i) on the halo scale radius. At fixed $M/\MBH$ the shift decreases rapidly with $a/\MBH$, as expected from Eq.~\eqref{firstorder}. For $a=\MBH$ the curves leave the Sgr\,A$^*$ band at $M/\MBH\simeq0.21$, $0.26$, and $0.32$ and the M87$^*$ band at $0.56$, $0.69$, and $0.83$ (for $\sigma=1/2$, $1$, and $3/2$). For $a=3\MBH$, only the $\sigma=1/2$ curve leaves the M87$^*$ band within $M\le\MBH$ (at $M\simeq0.87\,\MBH$; at $M=\MBH$, $\rSh/\MBH=6.45$, $6.21$, and $6.05$). For $a=10\,\MBH$ the curves exceed the Sgr\,A$^*$ band only for $M/\MBH\gtrsim0.76$-$0.95$, and for $a=30\,\MBH$ all curves remain inside both bands up to $M=\MBH$, with $\rSh/\MBH\le5.37$. The dependence on $\sigma$ amounts to a rescaling of $M$ by a factor $\lesssim1.5$ between $\sigma=1/2$ and $3/2$, and it fades as $a$ grows, in agreement with the asymptotic form of $I_\sigma$.
	
	Fig.~\ref{fig:Band}(b) reveals an important feature of convention (ii): the curves lie close to the dotted line $3\sqrt3\,\MBH/\MADM$, and the bounds in Table~\ref{Tab:bounds} depend only weakly on $a$ and $\sigma$. As $a$ grows, they approach $M/\MBH\simeq0.23$ (Sgr\,A$^*$) and $0.094$ (M87$^*$), the values at which a Schwarzschild shadow of mass $\MBH$ would be inconsistent with a dynamical mass $\MBH+M$. In other words, in convention (ii) the bounds are dominated by the mismatch between the mass that sets the shadow and the mass measured dynamically, and the lensing by the halo, which enlarges the shadow, only weakens them slightly. These bounds therefore constrain the amount of mass, of any nature, that can reside between the photon sphere and the radii probed by the mass measurements, rather than the details of the Hernquist profile.
	
	The bounds should be regarded as indicative. They inherit the systematic uncertainties of the mapping between the observed emission ring and the critical curve, as well as those of the mass and distance priors \cite{EventHorizonTelescope:2019ggy,EventHorizonTelescope:2022xqj}; moreover, we have neglected the BH spin, whose effect on the shadow size is small \cite{Vagnozzi:2022moj}. More importantly, they apply only to compact halos. For a galactic halo, $M/a\simeq4.8\times10^{-5}\,(M/10^{12}\Msun)(a/{\rm kpc})^{-1}$ in geometric units, i.e., $5\times10^{-6}$-$5\times10^{-5}$ for $M=10^{12}\Msun$ and $a=1$-$10$\,kpc. Equation~\eqref{firstorder} then gives $\delta\rSh/\rSh\sim M/a$, several orders of magnitude below the precision of current and planned horizon-scale imaging.
	
	\section{Quasinormal modes}\label{sec:QNM}
	
	\subsection{Master equation}\label{sec:master}
	
	We consider electromagnetic ($s=1$) and axial gravitational ($s=2$) perturbations of the background \eqref{met1}. The electromagnetic field is treated as a test field that does not interact with the DM. For the gravitational sector we restrict to axial perturbations: within the approach of Refs.~\cite{Cardoso:2021wlq,Cardoso:2022whc}, axial matter perturbations decouple from the metric perturbations, whereas polar gravitational perturbations couple to the fluid \cite{Speeney:2024mas} and are not treated here. Decomposing
	\begin{equation}
		\Psi_s(t,r,\theta,\phi)=e^{-i\omega t}\,\frac{\psi_s(r)}{r}\,Y_{\ell m}(\theta,\phi),
	\end{equation}
	where $Y_{\ell m}$ are spherical harmonics (tensor harmonics of axial parity for $s=2$) and $\ell$ is the multipole number, the radial function obeys
	\begin{equation}
		\frac{d^2\psi_s}{dr_*^2}+\left[\omega^2-V_{s,\ell}(r)\right]\psi_s=0,\label{eq:radial_equation}
	\end{equation}
	with the tortoise coordinate defined by
	\begin{equation}
		\frac{dr_*}{dr}=\left[A(r)\left(1-\frac{2m(r)}{r}\right)\right]^{-1/2}.\label{tortoise}
	\end{equation}
	For a metric of the form \eqref{genmet} with the closure \eqref{eq:Phi_prime}, the effective potentials are \cite{Cardoso:2021wlq,Cardoso:2022whc,Figueiredo:2023gas,Pezzella:2024tkf}
	\begin{equation}
		V_{s,\ell}=A(r)\left[\frac{\ell(\ell+1)}{r^2}+\frac{2m(1-s^2)}{r^3}-\frac{m'(1-s)}{r^2}\right],\label{Veffs}
	\end{equation}
	that is,
	\begin{align}
		V_{1,\ell}&=A(r)\,\frac{\ell(\ell+1)}{r^2},\label{EffPotEM}\\
		V_{2,\ell}&=A(r)\left[\frac{\ell(\ell+1)}{r^2}-\frac{6m(r)}{r^3}+\frac{m'(r)}{r^2}\right].\label{EffPot}
	\end{align}
	For $s=0$, Eq.~\eqref{Veffs} reproduces the potential of a massless scalar field, $A\ell(\ell+1)/r^2+(2r)^{-1}d[A(1-2m/r)]/dr$, as can be verified using the identity $\Phi'(1-2m/r)=2m/r^2$ implied by Eq.~\eqref{eq:Phi_prime}. For $m=\MBH$, Eqs.~\eqref{EffPotEM} and \eqref{EffPot} reduce to the Regge-Wheeler potentials \cite{Regge:1957td}. The electromagnetic potential is proportional to $A/r^2$ and therefore peaks exactly at the photon sphere, $r_0=\rPS$. For $s=2$ and $\ell=2$, the peak lies at $r_0\simeq3.28$-$3.31\,\MBH$ for the parameters considered here. Both lie outside the region $r<r_{\rm DEC}$ discussed in Sec.~\ref{sec:EC}.
	
	\subsection{WKB approximation}\label{sec:WKB}
	
	At sixth order, the WKB quantization condition reads \cite{Iyer:1986np,Konoplya:2003ii}
	\begin{equation}
		\frac{i\,(\omega^2-V_0)}{\sqrt{-2V_0''}}-\sum_{j=2}^{6}\Lambda_j=n+\frac12,\label{eq:wkb_condition}
	\end{equation}
	where $V_0=V_{s,\ell}(r_0)$ is the maximum of the potential, $V_0''=d^2V_{s,\ell}/dr_*^2|_{r_0}$, $n=0,1,\dots$ is the overtone number, and the corrections $\Lambda_j$ depend on $n$ and on the derivatives of $V_{s,\ell}$ with respect to $r_*$ up to order $2j$ at $r_0$. With our conventions, QNMs have $\omega_R>0$ and $\omega_I<0$. For $\sigma\ne1$, $\PhiH$ is obtained by numerical quadrature of Eq.~\eqref{PhiHprime}; the $r_*$-derivatives of $V_{s,\ell}$ are computed from the analytic $r$-derivatives of $\PhiH'$, $m$, and $f$, so that only the value $\PhiH(r_0)$ requires the numerical integral. The WKB expansion is asymptotic and is most accurate for $n<\ell$; we therefore restrict to $\ell=2$ and $n=0,1$. As an estimate of the error we use the relative difference between the sixth- and fifth-order results, $\varepsilon\equiv|\omega^{(6)}-\omega^{(5)}|/|\omega^{(6)}|$.
	
	\subsection{Eikonal correspondence}\label{sec:eik}
	
	For $\ell\gg1$, the QNMs of static, spherically symmetric BHs are determined by the unstable circular null geodesics \cite{Cardoso:2008bp}:
	\begin{equation}
		\omega_R\simeq\left(\ell+\tfrac12\right)\Omega_c,\qquad\Omega_c=\frac{1}{b_c},\qquad|\omega_I|\simeq\left(n+\tfrac12\right)\lambda,\label{eikonal}
	\end{equation}
	where $\Omega_c$ is the coordinate angular velocity of the circular photon orbit and $\lambda$ its Lyapunov exponent. For the metric \eqref{met1}, linearizing Eq.~\eqref{Veff} around $\rPS$ and using $\rPS\Phi'(\rPS)=2$ gives
	\begin{equation}
		\lambda^2=-\frac{A(\rPS)}{2}\left[1-\frac{2m(\rPS)}{\rPS}\right]\left[\frac{2}{\rPS^2}+\Phi''(\rPS)\right],\label{lyap}
	\end{equation}
	which reduces to $\lambda=1/(3\sqrt3\,\MBH)$ for Schwarzschild. Since the halo increases $b_c$ (Sec.~\ref{sec:sh}), Eq.~\eqref{eikonal} predicts a decrease of $\omega_R$. The sign of the shift of $|\omega_I|$ is set by that of $\lambda$, which we find numerically to decrease as well (Table~\ref{Tab:eik}). Equation~\eqref{eikonal} also implies that the quality factor $Q\equiv\omega_R/(2|\omega_I|)$ of the fundamental mode tends to $(\ell+\frac12)\Omega_c/\lambda$; being a ratio of frequencies, $Q$ does not depend on the mass used as a unit.
	
	\subsection{Results}\label{sec:QNMres}
	
\begin{figure*}[t]
		\centering
		\includegraphics[width=\textwidth]{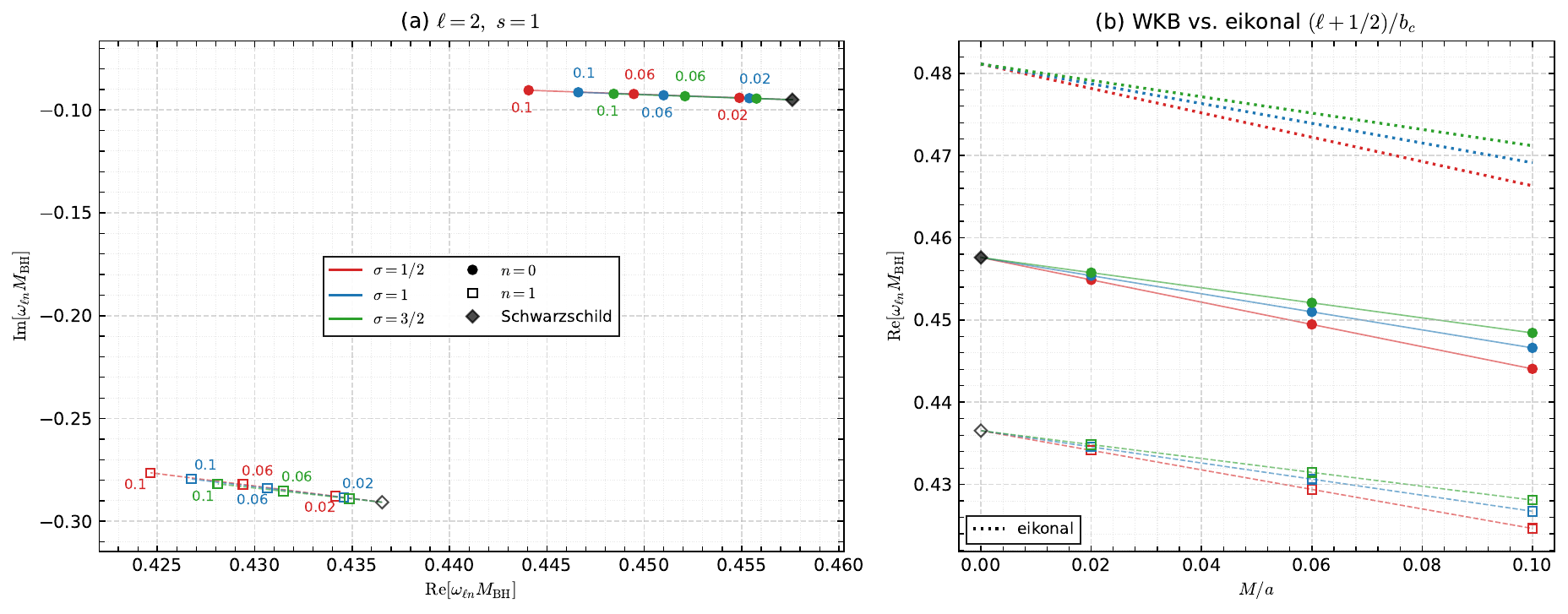}\\[2mm]
		\includegraphics[width=\textwidth]{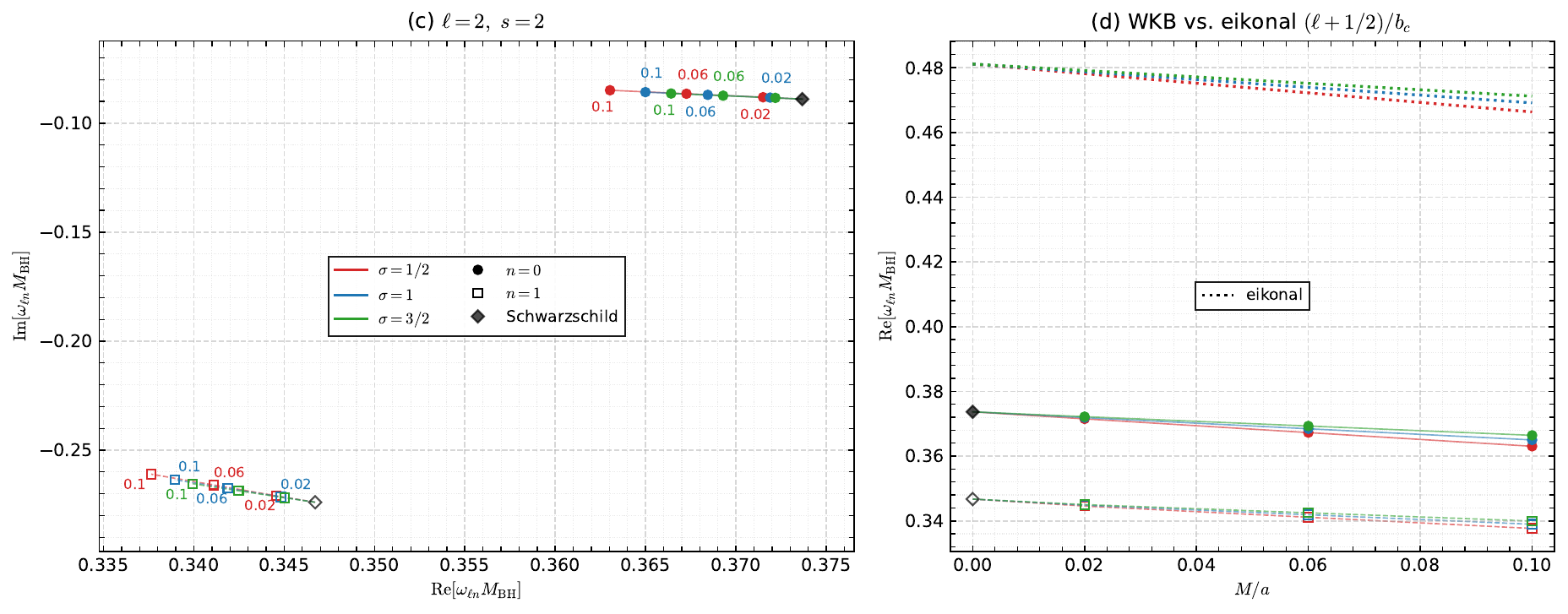}
		\caption{Sixth-order WKB QNM frequencies for $\ell=2$ and $a=\MBH$, for $M/a=0.02$, $0.06$, and $0.1$ (labels), $\sigma=1/2$ (red), $1$ (blue), and $3/2$ (green), and $n=0$ (filled symbols) and $n=1$ (open symbols). Top, $s=1$: (a) QNMs in the complex plane; (b) $\MBH\,\omega_R$ as a function of $M/a$, compared with the eikonal estimate $(\ell+\frac12)/b_c$ (dotted lines). Bottom, $s=2$: (c) and (d), same quantities. The gray diamonds are the exact Schwarzschild values ($M=0$).}
		\label{fig:WKB6s2}
	\end{figure*}

	The sixth-order WKB frequencies for $\ell=2$, $n=0,1$, and $a=\MBH$ are listed in Table~\ref{Tab:QNM} together with the error estimate $\varepsilon$, and are shown in Fig.~\ref{fig:WKB6s2}.
	
	\begin{table*}[t]
		\caption{QNM frequencies $\MBH\,\omega$ of electromagnetic ($s=1$) and axial gravitational ($s=2$) perturbations for $\ell=2$, $n=0,1$, and $a=\MBH$, obtained with the sixth-order WKB method. The columns $\varepsilon$ give the relative difference between the sixth- and fifth-order results, in percent. The rows $M=0$ give the Schwarzschild values obtained with the same method and, in brackets, the exact values. Fractional shifts with respect to the Schwarzschild WKB values are less affected by the WKB truncation error than the absolute frequencies.}
		\label{Tab:QNM}
		\begin{ruledtabular}
			\begin{tabular}{cccccc}
				$\sigma$ & $M/a$ & $\MBH\,\omega$ ($n=0$) & $\varepsilon$ (\%) & $\MBH\,\omega$ ($n=1$) & $\varepsilon$ (\%)\\
				\colrule
				\multicolumn{6}{c}{Electromagnetic perturbations, $s=1$}\\
				\colrule
				-- & 0 & $0.45759-0.09501i$ \;[$0.45759-0.09500i$] & 0.0025 & $0.43653-0.29073i$ \;[$0.43653-0.29073i$] & 0.00031\\
				1/2 & 0.02 & $0.45488-0.09408i$ & 0.0022 & $0.43415-0.28783i$ & 0.0012\\
				& 0.06 & $0.44946-0.09223i$ & 0.0015 & $0.42940-0.28207i$ & 0.0029\\
				& 0.10 & $0.44406-0.09040i$ & 0.00094 & $0.42465-0.27638i$ & 0.0046\\
				1   & 0.02 & $0.45539-0.09428i$ & 0.0021 & $0.43457-0.28845i$ & 0.0015\\
				& 0.06 & $0.45100-0.09283i$ & 0.0012 & $0.43065-0.28392i$ & 0.0038\\
				& 0.10 & $0.44661-0.09139i$ & 0.00039 & $0.42674-0.27941i$ & 0.0062\\
				3/2 & 0.02 & $0.45576-0.09444i$ & 0.0022 & $0.43485-0.28893i$ & 0.00080\\
				& 0.06 & $0.45209-0.09329i$ & 0.0015 & $0.43147-0.28536i$ & 0.0018\\
				& 0.10 & $0.44843-0.09215i$ & 0.00082 & $0.42810-0.28179i$ & 0.0028\\
				\colrule
				\multicolumn{6}{c}{Axial gravitational perturbations, $s=2$}\\
				\colrule
				-- & 0 & $0.37362-0.08889i$ \;[$0.37367-0.08896i$] & 0.031 & $0.34630-0.27348i$ \;[$0.34671-0.27391i$] & 0.17\\
				1/2 & 0.02 & $0.37150-0.08809i$ & 0.029 & $0.34456-0.27098i$ & 0.16\\
				& 0.06 & $0.36726-0.08649i$ & 0.025 & $0.34110-0.26600i$ & 0.15\\
				& 0.10 & $0.36303-0.08489i$ & 0.021 & $0.33764-0.26104i$ & 0.13\\
				1   & 0.02 & $0.37189-0.08826i$ & 0.030 & $0.34482-0.27149i$ & 0.17\\
				& 0.06 & $0.36844-0.08700i$ & 0.029 & $0.34188-0.26753i$ & 0.17\\
				& 0.10 & $0.36499-0.08575i$ & 0.028 & $0.33894-0.26359i$ & 0.16\\
				3/2 & 0.02 & $0.37218-0.08839i$ & 0.032 & $0.34502-0.27187i$ & 0.18\\
				& 0.06 & $0.36929-0.08738i$ & 0.034 & $0.34247-0.26865i$ & 0.19\\
				& 0.10 & $0.36641-0.08637i$ & 0.036 & $0.33993-0.26544i$ & 0.20\\
			\end{tabular}
		\end{ruledtabular}
	\end{table*}
	
	Figs.~\ref{fig:WKB6s2}(a) and \ref{fig:WKB6s2}(c) show that, as $M/a$ increases, the QNMs of each overtone move away from the Schwarzschild value along an almost straight line in the complex plane: $\omega_R$ decreases and $\omega_I$ becomes less negative. At fixed $\MBH$, the halo therefore lowers both the oscillation frequency and the damping rate, i.e., the ringdown is slower and longer lived. The three values of $\sigma$ produce displacements in nearly the same direction, and differ mainly in their magnitude, which is largest for $\sigma=1/2$ and smallest for $\sigma=3/2$; this is the same ordering as for the shadow (Sec.~\ref{sec:shres}). For $s=1$, $n=0$, and $M/a=0.1$, $\omega_R$ decreases by $2.96\%$, $2.40\%$, and $2.00\%$, and $|\omega_I|$ by $4.85\%$, $3.81\%$, and $3.01\%$, for $\sigma=1/2$, $1$, and $3/2$, respectively, relative to the Schwarzschild WKB values. The $n=1$ modes show comparable relative shifts; for instance, for $s=1$, $\sigma=1/2$, and $M/a=0.1$, $\omega_R$ decreases by $2.72\%$ and $|\omega_I|$ by $4.94\%$. The shifts are approximately linear in $M/a$ over the range considered.
	
	Figs.~\ref{fig:WKB6s2}(b) and \ref{fig:WKB6s2}(d) compare the WKB values of $\omega_R$ with the eikonal estimate $(\ell+\frac12)/b_c$. For $\ell=2$ the absolute values differ, as expected for the lowest radiative multipole: for Schwarzschild, the eikonal value $0.481/\MBH$ exceeds the exact frequency by $5\%$ for $s=1$ and by $29\%$ for $s=2$, for which the spin-dependent terms in Eq.~\eqref{EffPot} are not negligible at $\ell=2$. The eikonal relations nevertheless capture the dependence on the halo, i.e., the slopes of the curves. Table~\ref{Tab:eik} compares the fractional shifts of the $n=0$ modes at $M/a=0.1$ with the predictions $\delta\Omega_c/\Omega_c$ and $\delta\lambda/\lambda$ obtained from Eqs.~\eqref{sinxi} and \eqref{lyap}.
	
	\begin{table*}[t]
		\caption{Fractional shifts, in percent, of the $n=0$, $\ell=2$ QNMs at $M/a=0.1$ and $a=\MBH$, with respect to the Schwarzschild values computed with the same sixth-order WKB method, compared with the eikonal predictions. $Q=\omega_R/(2|\omega_I|)$ is the quality factor, whose eikonal limit is $(\ell+\frac12)\Omega_c/\lambda$.}
		\label{Tab:eik}
		\begin{ruledtabular}
			\begin{tabular}{cccccccccc}
				& \multicolumn{3}{c}{Eikonal} & \multicolumn{3}{c}{$s=1$ (WKB)} & \multicolumn{3}{c}{$s=2$ (WKB)}\\
				$\sigma$ & $\delta\Omega_c/\Omega_c$ & $\delta\lambda/\lambda$ & $\delta(\Omega_c/\lambda)/(\Omega_c/\lambda)$ & $\delta\omega_R/\omega_R$ & $\delta|\omega_I|/|\omega_I|$ & $\delta Q/Q$ & $\delta\omega_R/\omega_R$ & $\delta|\omega_I|/|\omega_I|$ & $\delta Q/Q$\\
				\colrule
				1/2 & $-3.08$ & $-4.93$ & $+1.95$ & $-2.96$ & $-4.85$ & $+1.99$ & $-2.83$ & $-4.50$ & $+1.74$\\
				1   & $-2.49$ & $-3.88$ & $+1.44$ & $-2.40$ & $-3.81$ & $+1.47$ & $-2.31$ & $-3.54$ & $+1.27$\\
				3/2 & $-2.07$ & $-3.05$ & $+1.01$ & $-2.00$ & $-3.01$ & $+1.04$ & $-1.93$ & $-2.83$ & $+0.93$\\
			\end{tabular}
		\end{ruledtabular}
	\end{table*}
	
	The eikonal formulas reproduce the WKB shifts of $\omega_R$ and $|\omega_I|$ to within $4\%$ of their size for $s=1$ and within $9\%$ for $s=2$, and they reproduce the shift of the quality factor of the electromagnetic modes to within $3\%$, even though $\ell=2$ is far from the eikonal regime. The better agreement for $s=1$ is expected, since the electromagnetic potential is proportional to $A/r^2$, whose maximum is the photon sphere itself, whereas for $s=2$ the terms $-6m/r^3+m'/r^2$ shift the peak and modify its curvature. This agreement provides an independent consistency check between the WKB calculation and the geodesic calculation of Sec.~\ref{sec:shadow}, and it explains why the QNMs and the shadow display the same ordering in $\sigma$. The Lyapunov exponent is about $1.5$-$1.6$ times more sensitive to the halo than $\Omega_c$, which is why the damping rate is more affected than the oscillation frequency, and why the quality factor increases. In the weak-halo limit, Eq.~\eqref{firstorder} gives $\delta\omega_R/\omega_R\simeq\delta\Omega_c/\Omega_c\simeq-M\,I_\sigma(a)$, i.e., $-3.09\%$, $-2.50\%$, and $-2.07\%$ for $M/a=0.1$, in agreement with the second column of Table~\ref{Tab:eik}.
	
	As for the shadow, the statements above refer to frequencies in units of $\MBH$. In units of the total mass, $\MADM\,\omega=(1+M/\MBH)\,\MBH\,\omega$. For $a=\MBH$ and $M/a=0.1$, the factor $1+M/\MBH=1.1$ overcompensates the halo-induced decrease: relative to the Schwarzschild value of the same total mass, $\omega_R$ of the fundamental mode is larger by $6.7\%$, $7.4\%$, and $7.8\%$ for $s=1$ and $\sigma=1/2$, $1$, and $3/2$ (and by $6.9\%$, $7.5\%$, and $7.9\%$ for $s=2$), and $|\omega_I|$ is larger by $4.7$-$6.9\%$. A ringdown analysis that ignores the halo would thus infer a mass that differs from both $\MBH$ and $\MADM$. Since a single QNM frequency is degenerate with the mass scale, a halo signature can only be identified through dimensionless combinations such as the quality factor, which increases by $0.9$-$2.0\%$ for $M/a=0.1$ (Table~\ref{Tab:eik}), or through the ratios of several modes. For rotating BHs, however, $Q$ also depends on the spin, and a shift of this size could be partially degenerate with a small spin; a dedicated analysis including rotation would be required to assess its detectability.
	
	The differences between $\sigma=1/2$, $1$, and $3/2$ are quantitative rather than qualitative. They originate entirely from the different exterior density profiles, since, within the WKB approximation, the QNMs depend only on the potential in the neighborhood of its peak at $r_0\simeq3.0$-$3.3\,\MBH$. In particular, they are not affected by the continuation of $\rho_1$ inside the horizon discussed in Sec.~\ref{sec:embedded}. The member $\sigma=1/2$, which has the largest halo mass at small radii, produces the largest shifts, but it should be recalled that it is singular at the horizon (Sec.~\ref{sec:regularity}). The regular members, $\sigma=1$ and $3/2$, give shifts that differ by $20$-$30\%$.
	
	For $s=1$, the fifth- to sixth-order differences are $\varepsilon\lesssim0.003\%$ for $n=0$ and $\lesssim0.007\%$ for $n=1$. For $s=2$, they are $0.02$-$0.04\%$ for $n=0$ and $0.13$-$0.20\%$ for $n=1$, consistent with the deviation of the Schwarzschild WKB values from the exact ones in Table~\ref{Tab:QNM} ($0.02\%$ and $0.13\%$, respectively). For $s=2$ and $n=1$, the halo-induced shifts at $M/a=0.02$ ($0.5\%$ in $\omega_R$ and $0.9\%$ in $|\omega_I|$) are only a few times larger than $\varepsilon$; however, the truncation error largely cancels in the fractional shifts, which vary smoothly with $M/a$, so the trends in Fig.~\ref{fig:WKB6s2} are robust. A third-order WKB calculation reproduces the fractional shifts of Table~\ref{Tab:eik} to within $0.05$ percentage points, and an independent implementation of the WKB series to arbitrary order, based on its equivalence with Rayleigh-Schr\"odinger perturbation theory for the inverted anharmonic oscillator \cite{Hatsuda:2019eoj}, reproduces all entries of Table~\ref{Tab:QNM} (Appendix~\ref{app:numerics}). A confirmation by time-domain evolution or by a direct frequency-domain integration would further strengthen the overtone results.
	
	\section{Conclusions}\label{sec:Conc}
	
	We have constructed a one-parameter family of static, spherically symmetric spacetimes describing a Schwarzschild BH embedded in a Hernquist DM halo, labeled by the exponent $\sigma>0$ that controls how the halo density is switched off at the horizon, $\rho\propto(r-2\MBH)^\sigma$. The density vanishes at and inside the horizon, the mass function satisfies $m(2\MBH)=\MBH$ and $m(\infty)=\MBH+M$ for every $\sigma$, and the metric follows from the Einstein-cluster closure. It is given by a single quadrature, and in closed form for $\sigma=1$, the member that coincides with the solution of Cardoso \textit{et al.}~\cite{Cardoso:2021wlq}. We have used this family to assess how the shadow and the QNMs depend on the halo parameters and on the near-horizon behavior of the matter distribution. Our main conclusions are as follows.
	
	The curvature invariants and the energy density measured by freely falling observers are finite at the horizon if and only if $\sigma\ge1$; for $\sigma<1$ the horizon carries a curvature singularity, and for $\sigma=1$ the curvature is finite but discontinuous across it. The null, weak, and strong energy conditions hold everywhere outside the horizon, whereas the dominant energy condition is violated in a layer $2\MBH<r<r_{\rm DEC}$, with $r_{\rm DEC}=5m(r_{\rm DEC})/2\simeq2.50$-$2.58\,\MBH$ for the configurations studied. This violation is a manifestation of a more general limitation of the Einstein-cluster closure, which requires superluminal tangential velocities inside the photon sphere and therefore admits a particle interpretation only for $r>\rPS$.
	
	Within the closure, the photon sphere satisfies $\rPS=3m(\rPS)$. At fixed $\MBH$, we proved that the shadow radius exceeds $3\sqrt3\,\MBH$ for any $M>0$ and increases monotonically with $M$, while it decreases with $a$ and $\sigma$. To first order, the fractional shift is $M\,I_\sigma(a)$ [Eq.~\eqref{firstorder}], equal to $M/(a+3\MBH)$ for $\sigma=1$; the dependence on $\sigma$ disappears as $(\MBH/a)\ln(a/\MBH)$ for extended halos. For $a=\MBH$ and $M/a=0.1$, the shadow is enlarged by $2.1$-$3.2\%$, whereas the photon sphere moves by only $0.4$-$1.1\%$.
	
	The comparison with the EHT shadow sizes of Sgr\,A$^*$ and M87$^*$ constrains only compact halos. For $a=\MBH$, identifying the mass prior with $\MBH$ gives $M/\MBH\lesssim0.21$-$0.32$ (Sgr\,A$^*$, $2\sigma$) and $\lesssim0.56$-$0.83$ (M87$^*$, $1\sigma$), while identifying it with $\MADM$, which is the appropriate choice when the mass is measured dynamically on scales much larger than $a$, gives $M/\MBH\lesssim0.32$-$0.41$ and $\lesssim0.12$-$0.14$, respectively. In the latter convention the bounds depend only weakly on $a$ and $\sigma$, and approach $0.23$ and $0.094$ for extended halos, because they are dominated by the mismatch between the mass that sets the shadow and the mass measured dynamically. For galactic halos, $M/a\sim10^{-5}$ and the effect on the shadow is far below observational reach.
	
	For electromagnetic and axial gravitational perturbations with $\ell=2$, the sixth-order WKB frequencies show that, at fixed $\MBH$, the halo decreases both the oscillation frequency and the damping rate, and increases the quality factor. For $a=\MBH$ and $M/a=0.1$, the fundamental frequency decreases by $1.9$-$3.0\%$, the damping rate by $2.8$-$4.9\%$, and the quality factor increases by $0.9$-$2.0\%$, depending on $s$ and $\sigma$. These shifts are reproduced to within $4\%$ ($s=1$) and $9\%$ ($s=2$) of their size by the eikonal relations involving the critical impact parameter and the Lyapunov exponent of the photon sphere, which provides a quantitative link between the shadow and the ringdown even at $\ell=2$. In units of the total mass, the frequencies instead increase. Since the frequency of a single mode is degenerate with the mass scale, the halo is imprinted on the ringdown only through dimensionless combinations such as the quality factor or ratios of different modes.
	
	Several simplifying assumptions should be relaxed in future work. (i) The halo is isolated and extends to infinity, and the Einstein-cluster closure is consistent with a particle interpretation only outside the photon sphere; models with radial pressure, or matter distributions truncated at or beyond the innermost stable circular orbit, would avoid the superluminal region. (ii) The BH is nonrotating; since the spin also affects the shadow size and the quality factor, rotating generalizations are needed to quantify possible degeneracies. (iii) Polar gravitational perturbations, which couple to the fluid \cite{Speeney:2024mas}, were not considered. (iv) The overtone frequencies should be confirmed with Pad\'e-resummed WKB \cite{Matyjasek:2017psv,Konoplya:2019hlu}, time-domain evolution, or direct integration. (v) The configuration $a\simeq\MBH$ is a toy model of a compact DM concentration rather than a galactic halo; realistic spike profiles \cite{Gondolo:1999ef,Sadeghian:2013laa}, combined with the constraints from stellar orbits around Sgr\,A$^*$ \cite{GRAVITY:2020gka}, would sharpen the astrophysical interpretation. (vi) A systematic scan of the QNMs in $a/\MBH$, analogous to Fig.~\ref{fig:Band}, is left for future work.

	\section*{Acknowledgements}
	S.F. is funded by the Conselleria de Innovación, Universidades, Ciencia y Sociedad Digital of the Generalitat Valenciana and the European Social Fund through a postdoctoral fellowship APOSTD 2025 (CIAPOS/2024/461).

	\bibliography{mybib}

\appendix

\section{The general-$\sigma$ family}\label{app:family}

Let $m_\sigma$ be given by Eq.~\eqref{msigma}, $m_\sigma=\MBH+Mr^2(a+r)^{-2}f^{\sigma+1}$, with $f=1-2\MBH/r$ and $f'=2\MBH/r^2$. Differentiating,
\begin{align}
	m_\sigma'&=Mf^{\sigma}\left[\frac{2ar\,f}{(a+r)^3}+\frac{2(\sigma+1)\MBH}{(a+r)^2}\right]\nonumber\\
	&=\frac{2Mar\,f^{\sigma}}{(a+r)^3}\left[f+(\sigma+1)\MBH\frac{a+r}{ar}\right],
\end{align}
so that $\rho_\sigma=m_\sigma'/(4\pi r^2)$ takes the form \eqref{rhosigma} with
\begin{align}
	A_\sigma&=1-\frac{2\MBH}{r}+(\sigma+1)\MBH\left(\frac1r+\frac1a\right)\nonumber\\
	&=1+(\sigma-1)\frac{\MBH}{r}+(\sigma+1)\frac{\MBH}{a},
\end{align}
as in Eq.~\eqref{Asigma}. By construction, $m_\sigma(2\MBH)=\MBH$ and $m_\sigma(\infty)=\MBH+M$. For $\sigma=1/2$, $A_{1/2}=1+\frac{\MBH}{2r}\left(\frac{3r}{a}-1\right)$; for $\sigma=1$, $A_1=(a+2\MBH)/a$; and for $\sigma=3/2$, $A_{3/2}=1+\frac{\MBH}{2r}+\frac{5\MBH}{2a}$, which reproduce Eqs.~\eqref{rho1/2}-\eqref{rho3/2}. The function $A_\sigma$ is positive for $\sigma>0$ and $r\ge2\MBH$: for $\sigma\ge1$ it satisfies $A_\sigma\ge1$, while for $0<\sigma<1$ it increases with $r$ from $A_\sigma(2\MBH)=\frac{1+\sigma}{2}+(\sigma+1)\frac{\MBH}{a}>0$.

Writing $1-2m_\sigma/r=f(1-D_\sigma)$, with $D_\sigma$ given by Eq.~\eqref{Dsigma}, Eq.~\eqref{eq:Phi_prime} and $\Phi_{\rm BH}'=2\MBH/(r^2f)$ give
\begin{align}
	\PhiH'&=\Phi'-\frac{2\MBH}{r^2f}=\frac{2}{r^2f}\left[\frac{m_\sigma}{1-D_\sigma}-\MBH\right]\nonumber\\
	&=\frac{2Mf^{\sigma-1}}{(a+r)^2(1-D_\sigma)},
\end{align}
which is Eq.~\eqref{PhiHprime}. To first order in $M$, $\PhiH'\simeq2Mf^{\sigma-1}/(a+r)^2=2\,\delta m/(r^2f^2)$, which leads to Eq.~\eqref{firstorder}.

The first-order shift of the photon sphere is $\delta\rPS=3\,\delta m(3\MBH)$, with $\delta m(3\MBH)=9M\MBH^2(a+3\MBH)^{-2}f(3\MBH)^{\sigma+1}$ and $f(3\MBH)=1/3$, which gives Eq.~\eqref{drPS}. For $a=\MBH$ and $M/a=0.01$, Eq.~\eqref{drPS} gives $\delta\rPS/\MBH=0.00325$, $0.00188$, and $0.00108$ for $\sigma=1/2$, $1$, and $3/2$, in agreement with Table~\ref{Tab:ShandPS}.

For $a\gg\MBH$, expanding $f^{\sigma-1}\simeq1-2(\sigma-1)\MBH/r$ in Eq.~\eqref{firstorder} gives, to leading order in $\MBH/a$,
\begin{equation}
	I_\sigma(a)\simeq\frac{1}{a+3\MBH}\left[1-2(\sigma-1)\frac{\MBH}{a}\ln\frac{a}{3\MBH}\right].
\end{equation}
Numerically, $(a+3\MBH)\,I_{1/2}=1.236$, $1.129$, $1.033$ and $(a+3\MBH)\,I_{3/2}=0.827$, $0.899$, $0.972$ for $a/\MBH=1$, $10$, $100$.

\section{Numerical methods and validation}\label{app:numerics}

The halo potential $\PhiH(r)$ was obtained by adaptive quadrature of Eq.~\eqref{PhiHprime}, splitting the integration range and mapping the tail to $u=1/r$. The photon sphere was obtained as the root of $r-3m(r)$, and $b_c$ from Eq.~\eqref{sinxi}. The WKB frequencies were computed from Eq.~\eqref{eq:wkb_condition} with the $r_*$-derivatives of the potential obtained from the analytic $r$-derivatives of $m$, $f$, and $\PhiH'$. The following checks were performed:
\begin{enumerate}
	\item The relations $m_\sigma'=4\pi r^2\rho_\sigma$, $m_\sigma(2\MBH)=\MBH$, $m_\sigma(\infty)=\MBH+M$, and Eq.~\eqref{PhiHprime} were verified symbolically for $\sigma=1/2$, $1$, $3/2$, and $2.3$; Eqs.~\eqref{Gtt}-\eqref{Kretsch} were verified symbolically for a general mass function.
	\item The root of $r-3m(r)$ agrees to machine precision with the root of Eq.~\eqref{PScond} computed directly from $A(r)$, and $b_c$ from Eq.~\eqref{sinxi} agrees with the direct maximization of $A/r^2$.
	\item For $\sigma=1$, the numerical quadrature of $\PhiH$ reproduces Eq.~\eqref{PhiH1}.
	\item For $M\to0$, $\rPS\to3\MBH$, $\rSh\to3\sqrt3\,\MBH$, $\lambda\to1/(3\sqrt3\,\MBH)$, and the WKB frequencies tend to the Schwarzschild values listed in Table~\ref{Tab:QNM}.
	\item The first-order formula \eqref{firstorder} agrees with the full numerical shift of $\rSh$ to within $0.2$-$0.3\%$ of its value for $M=0.01\,\MBH$ and to within $1.9$-$2.7\%$ for $M=0.1\,\MBH$, for all three values of $\sigma$.
	\item The inequalities $\rPS>3\MBH$, $\rSh>3\sqrt3\,\MBH$, and $\PhiH<0$ hold for all configurations, and $r-3m(r)$ has a single root for $r>2\MBH$.
	\item The near-horizon behavior \eqref{Rnear} was verified symbolically, and the energy-condition combinations of Fig.~\ref{fig:EC01} were computed from $\rho=m'/(4\pi r^2)$ and Eq.~\eqref{eq:pt}.
	\item A third-order WKB calculation, with numerical $r_*$-derivatives of the potential, gives the Schwarzschild values $0.45713-0.09507i$ ($s=1$) and $0.37316-0.08922i$ ($s=2$) for $\ell=2$, $n=0$, in agreement with Ref.~\cite{Iyer:1986np}, and, e.g., $0.44368-0.09046i$ for $s=1$, $\sigma=1/2$, and $M/a=0.1$; the corresponding fractional shifts agree with those of Table~\ref{Tab:eik} to within $0.04$ percentage points.
	\item The WKB series was also computed to arbitrary order with an independent code that exploits the equivalence between the WKB expansion around the peak of the potential and Rayleigh-Schr\"odinger perturbation theory for the inverted anharmonic oscillator \cite{Hatsuda:2019eoj}, with the Taylor coefficients of $V_{s,\ell}(r_*)$ obtained by exact power-series arithmetic. This code reproduces all entries of Table~\ref{Tab:QNM}, including the error estimates $\varepsilon$, to the quoted precision.
\end{enumerate}

\end{document}